%% file: np_4_hrtf.tex
\documentclass[journal]{IEEEtran}

\ifCLASSINFOpdf
  \usepackage[pdftex]{graphicx}
  \graphicspath{{./figures/}}%
  \DeclareGraphicsExtensions{.pdf,.jpeg,.png}
\else
\fi

\usepackage{amsfonts}
\usepackage{bm}
\usepackage{dsfont}
\usepackage{relsize}
\usepackage{upgreek}
\usepackage{amsmath}
\usepackage{url}

\usepackage{xcolor}

\input{./utilities/math_commands.tex}

\begin{document}
\title{HRTF Interpolation using a Spherical Neural Process Meta-Learner}

\author{Etienne~Thuillier,~\IEEEmembership{Member,~IEEE,}
		and Craig~Jin~\IEEEmembership{Senior Member,~IEEE,}
        and Vesa~V\"{a}lim\"{a}ki,~\IEEEmembership{Fellow,~IEEE}%
\thanks{Manuscript received Oct. 18, 2023; revised xxx xx, xxxx.}%
\thanks{E.~Thuillier and V.~V\"{a}lim\"{a}ki are with the Acoustics Lab, Department
of Information and Communications Engineering, Aalto University, FI-02150 Espoo, Finland, e-mail: etienne.thuillier@aalto.fi.}%
\thanks{C.~T. Jin is with the School of Electrical and Computer Engineering, The University of Sydney, Sydney, Australia.}}%

\markboth{\tiny This work has been submitted to the IEEE/ACM T-ASL,~Oct.~2023, for possible publication. Copyright may be transferred without notice, after which this version may no longer be accessible. }%
{****Shell \MakeLowercase{\textit{et al.}}: Bare Demo of IEEEtran.cls for IEEE Journals}

\maketitle

\begin{abstract}
Several individualization methods have recently been proposed to estimate a subject's Head-Related Transfer Function (HRTF) using convenient input modalities such as anthropometric measurements or pinnae photographs. There exists a need for adaptively correcting the estimation error committed by such methods using a few data point samples from the subject's HRTF, acquired using acoustic measurements or perceptual feedback. To this end, we introduce a Convolutional Conditional Neural Process meta-learner specialized in HRTF error interpolation. In particular, the model includes a Spherical Convolutional Neural Network component to accommodate the spherical geometry of HRTF data. It also exploits potential symmetries between the HRTF's left and right channels about the median axis. In this work, we evaluate the proposed model's performance purely on time-aligned spectrum interpolation grounds under a simplified setup where a generic population-mean HRTF forms the initial estimates prior to corrections instead of individualized ones. The trained model achieves up to 3 dB relative error reduction compared to state-of-the-art interpolation methods despite being trained using only 85 subjects. This improvement translates up to nearly a halving of the data point count required to achieve comparable accuracy, in particular from 50 to 28 points to reach an average of -20 dB relative error per interpolated feature. Moreover, we show that the trained model provides well-calibrated uncertainty estimates. Accordingly, such estimates can inform the sequential decision problem of acquiring as few correcting HRTF data points as needed to meet a desired level of HRTF individualization accuracy.
\end{abstract}


\IEEEpeerreviewmaketitle

\section{Introduction}
\label{sec:introduction}

\input{./sections/introduction.tex}

\section{Background}
\label{sec:background}
\input{./sections/background.tex}

\section{Novel Model and Method}
\label{sec:model_and_method}
\input{./sections/model_and_method.tex}

\section{Results}
\label{sec:results}
\input{./sections/results.tex}

\section{Conclusion}
\label{sec:conclusion}
\input{./sections/conclusion.tex}

\ifCLASSOPTIONcaptionsoff
  \newpage
\fi

\bibliographystyle{IEEEtran}
\bibliography{IEEEabrv,./np_4_hrtf}

\begin{IEEEbiography}[{\includegraphics[width=1in,height=1.25in,clip,keepaspectratio]{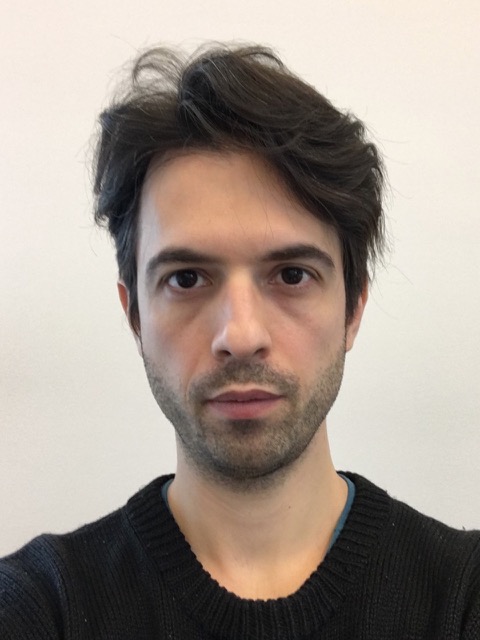}}]{Etienne Thuillier} is a PhD student from Aalto University (Espoo, Finland). His research lies in the application of machine learning for HRTF individualization. Originally a telecommunications engineer from Ecole Polytechnique de Montr\'eal, Etienne Thuillier started his career as a patent consultant before re-converting to technical practice through a startup creation experience and master's studies in audio signal processing and acoustics at Aalto University. He subsequently worked on research and development projects at Apple, Microsoft Research and Meta Reality Labs. %
\end{IEEEbiography}

\begin{IEEEbiography}[{\includegraphics[width=1in,height=1.25in,clip,keepaspectratio]{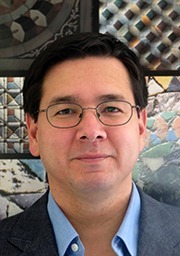}}]{Craig~T~.~Jin}
(M'92–SM'09) received the M.S. degree in applied physics from Caltech, Pasadena, California, in 1991 and the Ph.D. in electrical engineering from the University of Sydney, Sydney, Australia, in 2001. He is an Associate Professor in the School of Electrical and Computer Engineering, University of Sydney and Director of the Computing and Audio Research Laboratory. His research interest includes experimental and theoretical aspects of acoustic and biomedical signal processing. He has authored more than 200 papers in these research areas, holds eight
patents, and founded three start-up companies. He received national recognition in Australia (April 2005, Science in Public Fresh Innovators Program) for his invention of a spatial hearing aid.
\end{IEEEbiography}

\begin{IEEEbiography}[{\includegraphics[width=1in,height=1.25in,clip,keepaspectratio]{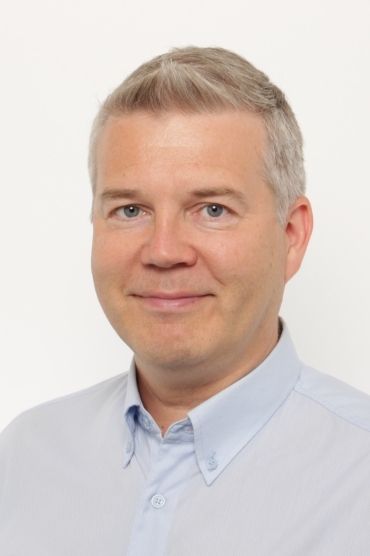}}]{Vesa V\"{a}lim\"{a}ki}
is a Full Professor of audio signal processing and Vice Dean of Research at Aalto University, Espoo, Finland. He received his MSc in Technology and Doctor of Science in Technology degrees, both in electrical engineering, from the Helsinki University of Technology (TKK) in 1992 and 1995, respectively. His doctoral dissertation dealt with fractional delay filters and physical modeling of musical instruments. 

In 1996 he was a Postdoctoral Research Fellow at the University of Westminster, London, UK. In 2001-2002 he was Professor of signal processing at the Pori School of Technology and Economics, Tampere University of Technology, Pori, Finland. In 2002, he was appointed Professor of audio signal processing at TKK. In 2008-2009, he was on sabbatical as a Visiting Scholar at the Center for Computer Research in Music and Acoustics (CCRMA), Stanford University, Stanford, CA. His research interests are in signal processing and machine learning applied to audio and music technology.

Prof.~V\"{a}lim\"{a}ki is a Fellow of the IEEE, a Fellow of the AES, and a Life Member of the Acoustical Society of Finland. In 2015-2020, he served a Senior Area Editor of the IEEE/ACM Transactions on Audio, Speech and Language Processing. He was the General Chair of the 14th Sound and Music Computing Conference SMC-17 in 2017. He is the Editor-in-Chief of the \emph{Journal of the Audio Engineering Society}.
\end{IEEEbiography}

\end{document}

%% file: utilities/math_commands.tex
\def\eqref#1{equation~\ref{#1}}

\def\1{\bm{1}}

\DeclareMathAlphabet{\mathsfit}{\encodingdefault}{\sfdefault}{m}{sl}
\SetMathAlphabet{\mathsfit}{bold}{\encodingdefault}{\sfdefault}{bx}{n}

\def\gC{{\mathcal{C}}}

\def\gG{{\mathcal{G}}}

\def\gM{{\mathcal{M}}}
\def\gN{{\mathcal{N}}}

\def\gS{{\mathcal{S}}}
\def\gT{{\mathcal{T}}}

\def\gX{{\mathcal{X}}}

\def\sR{{\mathbb{R}}}

\usepackage{cancel}

%% file: sections/introduction.tex
\IEEEPARstart{R}{ecent} adoption of augmented and virtual reality interfaces has pushed the need for immersive spatial audio rendering solutions that scale to mass market \cite{crum2019hearables, gupta2022armr, yang2022aar, herre2023mpeg}. The Head Related Transfer Function (HRTF) is a key component of current systems: it simulates the effect of the subject's body on the acoustic transmission channels between the subject's ears and sound sources as a function of their locations around the subject \cite{moller1996binaural}. Crucially, the HRTF is a function of the subject's morphology and is specific to each individual. Studies have shown that spatial audio percepts deteriorate when a generic HRTF is used for all subjects in a population compared to using individualized HRTF estimates~\cite{moller1996binaural, guezenoc2018hrtf}.
In this work, we propose the first HRTF interpolation method that provides well-calibrated uncertainty estimates. The method also demonstrates significantly improved interpolation accuracy with regards to the state of the art.

\subsection{Prior Art}

\par A recent review paper classifies HRTF individualization techniques into four categories defined by the source of HRTF information: acoustic measurements, numerical simulation, anthropometric data, and perceptual feedback~\cite{guezenoc2018hrtf}. As an alternative approach useful to our discussion, we classify below individualization techniques into two broad classes according to the way in which the subject's individualized HRTF is represented.

\par A first class of methods represents the subject's individualized HRTF in a non-parametric fashion with a sparse set of observed HRTF data points. In such approaches, interpolation methods are applied downstream to provide HRTF filter estimates at specified directions of arrival between the observed locations. Typically, the observations are collected using acoustic measurements~\cite{majdak2007multiple}, but recommender systems have also been proposed for composing the sparse set with HRTF filters derived from a pre-existing database and according to perceptual feedback obtained from the user~\cite{luo2015gaussian}.

\par Improvements in interpolation methods result in sparser set of observations becoming sufficient for meeting a required accuracy threshold, thereby accelerating the individualized HRTF acquisition process. Early methods include barycentric interpolation~\cite{sundareswara2003extensible}, natural neighbour interpolation~\cite{porschmann2020comparison}, spherical harmonic~\cite{zotkin2009regularized}, thin-plate spherical spline interpolation~\cite{carlile2002performance} and Gaussian process regression~\cite{luo2013kernel}. More recently, pre-processing has been shown to significantly reduce the required density of HRTF measurements needed to meet a given interpolation accuracy requirement~\cite{ben2011acoustic, aussal2014methodes, richter2014spherical, ben2019efficient}. Neural-network regressor models have also been proposed~\cite{ito2022head, lee2023global},  including a spherical convolutional neural network performing interpolation from a relatively dense equiangular grid counting 120 data points~\cite{chen2023headrelated}. Related works includes HRTF upsampling approaches using generative models~\cite{hogg2023hrtf}. However, such models currently provide improvements in the sparsest regimes only.%

\par A second class of methods parametrizes individualized HRTFs using low-dimensional latent-space representations embodied by fixed-length vectors of adjustable coefficients. Various approaches have been proposed to predict the coefficients of the representation including use of anthropometric measurement~\cite{zhi2022towards,yao2022individualization}, pinnae photographs~\cite{lee2018personalized}, HRTF observations~\cite{ito2022head}, perceptual feedback~\cite{yamamoto2017fully} or combinations thereof~\cite{miccini2021hybrid}. This provide a convenient means for promptly estimating a subject's HRTF from one or several input modalities. However, a common design compromise facing techniques in this class lies in providing a representation that is compact enough that prediction is facilitated, while retaining sufficient expressiveness that HRTF variability across the population is faithfully represented.
Due to the fixed dimensionality of latent representations in particular, and unlike the non-parametric interpolation methods described above, the expressiveness of the model does not scale with additional data points provided to it. More importantly, any resulting change to the HRTF representation is in this case global, such that any resulting local improvement is susceptible to adversely affect the representation elsewhere. This contrasts with non-parametric cases which provide representations that are local by construction.

\subsection{Problem}

\par There is a need for adaptively refining the individualized HRTF estimate provided by a parametric method until a pre-defined criterion of suitability is achieved, for example a user performance metric threshold under a listening test experiment. To this end,
we advocate for a hybrid approach to HRTF individualization in which the parametric estimate is corrected by integrating a few observations of the subject's HRTF using an interpolation method. Under this approach, the HRTF refinement problem can be framed as a sequential decision problem: that of acquiring as few correcting HRTF data points as needed to meet the performance requirement, using measurements or perceptual feedback. Such problem would benefit from using an accurate interpolation method that also provides well-calibrated uncertainty estimates. When suitably calibrated, uncertainty estimates can indeed be used to inform the choice of the next location to observe. Under a perceptual feedback acquisition scheme, they can additionally inform the selection of proposal HRTF filters to be submitted as queries to the subject. Finally, there also exists a need within augmented reality settings, for matching the rendered sound field with the user's surrounding acoustic environment. Such a problem could also be addressed using the suggested approach by adaptively refining the Binaural Room Transfer Function instead of the HRTF.

\subsection{Solution}

\par To facilitate the approach mentioned above, we introduce a novel model that we name Spherical Convolutional Conditional Neural Process (SConvCNP). The proposed model is a Convolutional Conditional Neural Process (ConvCNP) meta-learner~\cite{gordon2019convolutional} specialized in HRTF error interpolation. The model accommodates the spherical geometry of HRTF data. To this end, it includes a Spherical Convolutional Neural Network component~\cite{esteves2018learning,cohen2018spherical} which executes rotation-equivariant feature transforms. It also exploits the approximate symmetry between the HRTF’s left and right channels about the median axis. To the authors' best knowledge, this work is the first application of a Neural Process model to spherical data.%

\par Such a model learns a functional representation of the set of observed HRTF data points that preserves spatial structure and can be addressed at any location on the unit sphere. Furthermore, the representation is learned using rotation equivariant mappings which ensures the same transformation is applied with shared parameters everywhere on the sphere, irrespective of feature location. These aspects allow for learning local interpolations of the HRTF features in an sample-effective fashion. Moreover, the possibility, afforded by the model, to address any location on the unit sphere provides native compatibility for training on any HRTF databases irrespective of its data point grid layout.

\par This work implements and tests the SConvCNP model in a simplified experimental setup. Firstly, the interpolation is applied on the HRTF spectrum after time-alignment~\cite{ben2011acoustic, richter2014spherical, ben2019efficient}, leaving pure delay interpolation as future work for brevity. Secondly, a generic population-mean time-aligned spectrum is used as generic estimate for all subjects before correction, instead of individualized time-aligned spectra. This allows to evaluate the merits of the model purely from an interpolation performance standpoint, leaving the application to individualized HRTF correction as future work. The model is shown to achieve up to 3~dB of relative error reduction compared to state-of-the-art interpolation methods. This translates to nearly a halving of the required data to achieve a comparable level of accuracy. Moreover, our model is shown to provide well-calibrated uncertainty estimates.%

This paper is organized as follows. Sec.~\ref{sec:background} provides background on the ConvCNP model and its meta-training procedure. Sec.~\ref{sec:model_and_method} introduces the SConvCNP model, defines the interpolation tasks on which the model is trained, and proposes baseline and metrics for evaluating the model's performance both in terms of interpolation accuracy and uncertainty calibration. Sec.~\ref{sec:results} presents and discusses the experimental results. Sec.~\ref{sec:conclusion} concludes this paper.

%% file: sections/background.tex
\par In this section, we provide a technical review of the ConvCNP model and its meta-training procedure as background for the introduction of the SConvCNP model in Sec.~\ref{sec:model_and_method}.

\subsection{ConvCNP Architecture}
\label{sec:convcnp}

\par Neural Processes form a class of deep neural networks operating on sets to model stochastic processes~\cite{garnelo2018conditional,garnelo2018neural}. In neural process models, a set of observed location-feature data point pairs $\left\{\left(x_c,y_c\right)\right\}_{c=1}^C$ at the input informs a predictive distribution provided at the output for unseen values $y_t$ at target locations $x_t$, much in the same fashion as in Gaussian Processes~\cite{rasmussen2006gaussian}. In particular, the elements of the input set are subsumed into a representation embedding, which allows for handling  sets of different sizes and ensures invariance in the ordering of set elements~\cite{zaheer2017deepsets}. Recently, functional representation embeddings have been proposed that preserve the spatial structure in the input set and are addressable at any location coordinates $x_t$. These functional embeddings enable constructing translation equivariant neural process models, as appropriate when modeling stationary data~\cite{gordon2019convolutional, foong2020meta, dubois2020npf}. The ConvCNP is an example of such model~\cite{gordon2019convolutional}. A description of its architecture is given in the current section.

\begin{figure}[!t]
\centering
	\includegraphics[width=2.5in, clip]{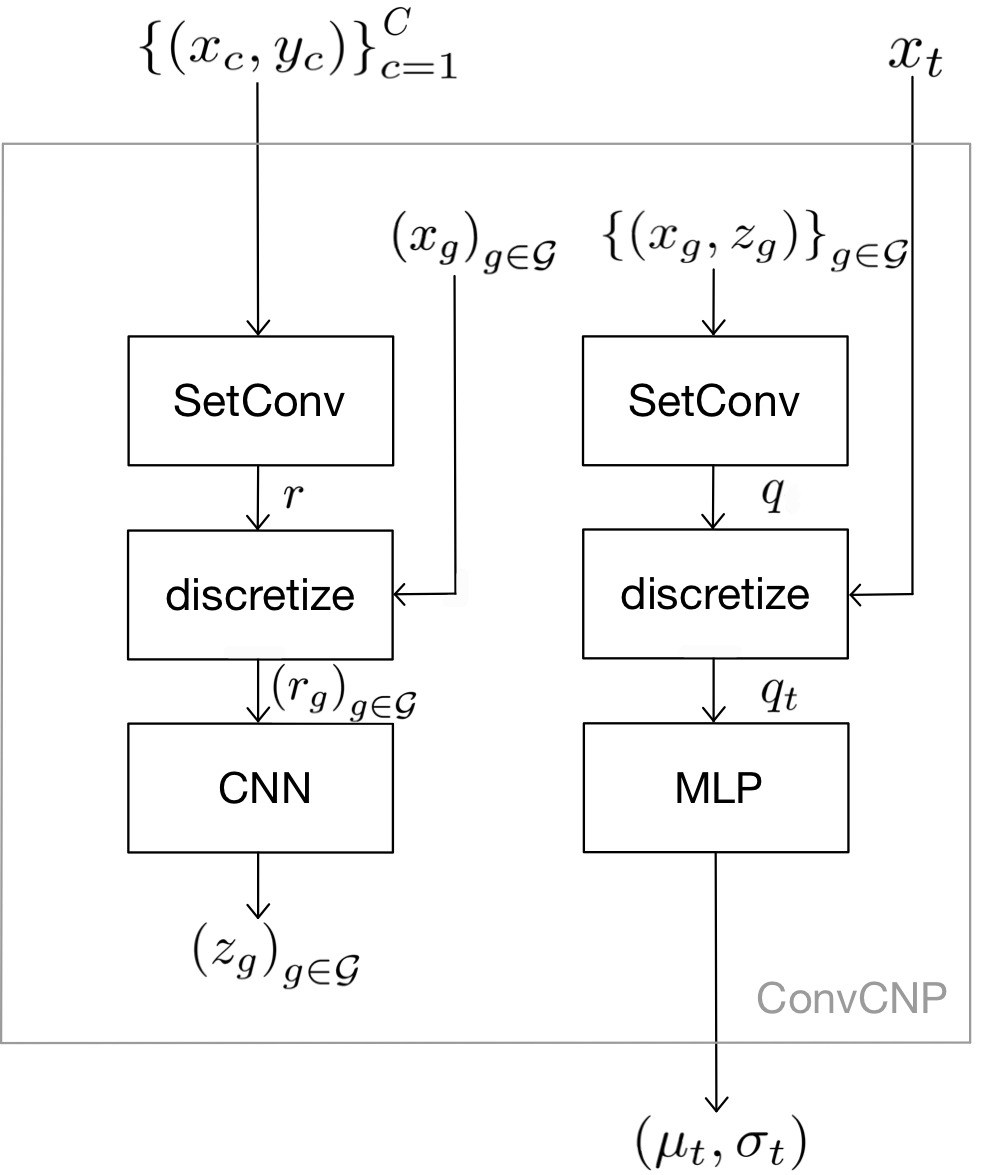}
\caption{Schematic block diagram of a typical ConvCNP model's architecture.\label{fig:encoder_decoder}}
\end{figure} %
\par A typical example of ConvCNP model architecture is given in the block diagram of Fig.~\ref{fig:encoder_decoder}. The model includes a first set convolution (block \emph{SetConv}) which maps a set of observed data points
$\left\{(x_c, y_c)\right\}_{c=1}^C$ into a functional representation~\cite{gordon2019convolutional,dubois2020npf}
\begin{align}
r &= \text{SetConv}\left(\left\{(x_c, y_c)\right\}_{c=1}^C\right),\label{eq:set_convolution_1}
\end{align}
which, assuming a multiplicity of one for data set elements~\cite{gordon2019convolutional}, returns a vector-valued point-wise representation
\begin{align}
	r(x) &=\left(\sum_{c=1}^C K(x_c,x),\ 
	\frac{\sum_{c=1}^C y_c K(x_c,x)}{\sum_{c=1}^C K(x_c,x)}\right),\label{eq:set_convolution_2}
\end{align}
at any specified location $x$. In the above expression, $K:\gX\times\gX \rightarrow \mathds{R}$ denotes a positive definite kernel with learnable parameter(s), for example, a Gaussian kernel in the case of planar data such as images~\cite{fasshauer2011positive}. Accordingly, the first channel of functional embedding $r$ before discretization, forms a kernel density of the observed locations: the result of a convolution between filter $K(x_c,\cdot)$ and a sum of unit-weighted Dirac distributions centered at locations $\left\{x_c\right\}_{c=1}^C$. The second channel forms an interpolant of the observed data points $\left\{(x_c,y_c)\right\}_{c=1}^C$ following the Nadaraya-Watson kernel regression method~\cite{nadaraya1964estimating}.

\par In ConvCNP models, translation equivariant representation learning is performed downstream of the set convolution using a Convolutional Neural Network (CNN). As pictured in Fig.~\ref{fig:encoder_decoder}, representation $r$ is first discretized following a grid $(x_g)_{g\in\gG}$ of regularly-spaced coordinates. Assuming two-dimensional planar data for example: $\gG=\left\{1,\hdots,G\right\}\times\left\{1,\hdots,G\right\}$, in which $G$ denotes the number of samples of the grid in each dimension. A second set convolution converts---at least implicitly---the learned representation at the output of the CNN back to a functional one~\cite{dubois2020npf}, denoted $q$ in the diagram. Crucially, $q$'s second channel\footnote{The first channel, representative of density, is less informative at this stage than in the case of the first set convolution and can optionally be discarded.} forms an interpolant of the learned representation $\left\{\left(x_g, z_g\right)\right\}_{g\in \gG}$ following (\ref{eq:set_convolution_2}).

\par Given the above, $q$ forms a learned functional representation of the input set which is spatially-structured and addressable at any user-specified target location $x_t$. In the example of Fig.~\ref{fig:encoder_decoder}, the resulting point-wise representation $q_t$ is decoded using a feed-forward neural network (MLP) decoder, which maps the target location $x_t$
to mean $\mu_t$ and standard deviation $\sigma_t$ values specifying a predictive distribution for the target features $y_t$ at that location:
\begin{align*}
	p\left(y_t\ \left|\ x_t,\ \left\{(x_c,y_c) \right\}_{c=1}^C\right.\right) %
	&\approx \gN\left(y_t;\ \mu_t, \sigma_t^2\right),
\end{align*}
where we assume uni-variate features for simplicity and $\gN$ denotes the normal distribution.

\subsection{Meta-training}

\par Features and advantages of the model are best understood under the lens of the meta-learning framework~\cite{schmidhuber1987evolutionary}. Under this perspective, the observed data-points $\left\{(x_c, y_c)\right\}_{c=1}^C$ form a task-specific train set and the ConvCNP model forms a meta-learning algorithm mapping the set to a trained discriminator $\text{MLP}\circ q$
that, given query location $x_t$, returns a predictive distribution $\gN(y_t;\mu_t,\sigma_t^2)$.  In particular, the learned functional embedding $q$ of train set $\left\{(x_c,y_c)\right\}_{c=1}^C$ parametrizes said discriminator MLP$\circ q$ such that the set's data point locations $\left\{x_c\right\}_{c=1}^C$ can be leveraged to provide well-calibrated uncertainty estimates $\sigma_t$. This is unlike common learning methods which generally discard such information and, as a consequence, provide trained models that cannot recognize when queried far from the points of the train set. While it remains possible for these models to quantify the uncertainty resulting from noise present in the labels (``data uncertainty'', ``aleatoric uncertainty''), the uncertainty in the choice of model and the value of its parameters (``model uncertainty'', ``epistemic uncertainty''), in particular as it relates to the train set used to optimize parameter values, is typically unaccounted for. In contrast, ConvCNP models are shown to provide well-behaved uncertainty estimates for stationary data~\cite{gordon2019convolutional}. This results in part from translation equivariance. Indeed, this ensures the model's outputs can be computed as function of distance to the context set's data points but not as a function of the data points' coordinate values themselves.

\par Model optimization within the meta-learning framework is carried out on a set of learning tasks (the meta-training set), each defined by a context (train) set and target (test) set pair. In particular, ConvCNP models can be trained following the maximum-likelihood objective~\cite{garnelo2018conditional}
\begin{equation}
	\max_\theta \sum_{(\gC, \gT) \in \gM}\ \sum_{(x, y)\in \gT} \log p(y \left|x,\gC;\ \theta\right), \label{eq:objective}
\end{equation}
where $\theta$ denotes the coefficient vector of the model's learnable parameters, $\gM=\left\{\left(\gC_m, \gT_m\right)\right\}_{m=1}^M$ denotes the meta-training set, $\gC_m=\{(x_c, y_c)\}_{c=1}^C$ forms a context (train) set, and $\gT_m=\{(x_t, y_t)\}_{t=1}^T$ forms a target (test) set. Additional validation and test meta-sets composed of held-out data are used to perform model selection and evaluate generalization performance.

%% file: sections/model_and_method.tex
\par In this section, we introduce the SConvCNP model and define the interpolation tasks on which the model is trained. Furthermore, we select the baselines and metrics for the purpose of evaluating the model's performance both in terms of interpolation accuracy and uncertainty calibration.

\subsection{Interpolation Task}
\label{sec:hrtf_interpolation_task}

\par When applied to the problem of HRTF interpolation, each task $(\gC,\gT)$ composing the meta-training set $\gM$ consists in interpolating a given subject's HRTF to specified unseen (target, test) locations given a set of observed (context, train) HRTF data points acquired from the subject. A detailed description of the specific interpolation task studied in this work follows.

\par Consider the following time-alignment factorization of the HRTF spectrum~\cite{ben2011acoustic, richter2014spherical, ben2019efficient}:
\begin{align}
	h(x) &= \left(\text{e}^{\text{i}  \frac{2\pi n}{N} \tau(x)} \right)_{n=0}^{N/2}\odot m(x),\label{eq:hrtf}
\end{align}
where
\begin{itemize}
	\item $\odot$ denotes the Hadamard (element-wise) product,
	\item $x \in \gS^2=\left\{x\in\mathds{R}^3\ \left|\ \|x\|_2=1\right.\right\}$ denotes the sound source direction represented in cartesian coordinates on the unit sphere,
	\item N denotes the filter tap count of the Head-Related Impulse Response (HRIR),
	\item $\tau: \gS^2\rightarrow [0,\infty)^2$ returns the pure delay values for both ears at specified location $x$,
	\item $m: \gS^2 \rightarrow \mathds{C}^{\left(N/2 + 1\right)\times 2}$ returns the positive frequency side of the time-aligned HRTF spectrum for both ears at specified location $x$,
	\item $h: \gS^2\rightarrow \mathds{C}^{\left(N/2 + 1\right)\times 2}$ returns the positive frequency side of the HRTF spectrum for both ears at specified location $x$.
\end{itemize}
\par Under this factorization, the time-aligned spectrum is composed of the minimum-phase and nonlinear phase all-pass components of the HRTF~\cite{nam2008method}.

\par Inspection of (\ref{eq:hrtf}) reveals that interpolating the pure delay $\tau$ and the time-aligned spectrum $m$ is in principle less challenging than interpolating spectrum $h$ directly. Indeed, the exponential factor in Equation (\ref{eq:hrtf}) maps pure delay values on the unit sphere to complex values, which real and imaginary parts ripple on the surface of $\gS^2$ following the spatial variations of pure delay $\tau$, at a rate proportional to normalized frequency $n/N$. Consequently, this exponential factor significantly contributes to the irregularity of the HRTF spectrum, especially in the higher portion of the frequency range. In effect, pure delay and aligned spectrum components have been shown to require spherical harmonic representations of greatly reduced order compared to the non-processed spectra for comparable reconstruction accuracy~\cite{ben2011acoustic, richter2014spherical, ben2019efficient, porschmann2020method}.

\par In this work, we employ simulated HRTFs from the HUTUBS database without changes to its coordinate system, which places the origin at the center of the subject's head~\cite{brinkmann2019a}.
We extract the time-aligned spectrum $m$ by factoring out the pure-delay exponential term out of (\ref{eq:hrtf}) for each data point of the HRTF set individually.
In particular, the pure delay is estimated in a preliminary step as the power-weighted average of excess group delay~\cite{nam2008method}.
More specifically, the weighted-average is computed using frequency bins lying within the 0 to 1.1 kHz frequency range.
This avoids sharp group delay jumps occurring around zeros of the HRTF spectrum in the upper frequency range~\cite{nam2008method}.
When applied to the simulated HRTFs of the HUTUBS database, this approach provides pure delay values that are spatially smooth.
We apply this time-alignment method to down-sampled versions of the binaural filters from 44.1 to 33.075 kHz.
This reduces the HRIR tap count from $N=256$ to $N=192$, thereby lowering memory requirements for running the model.

\par For brevity, we limit the experiments of this work to the interpolation of the time-aligned spectrum $m$ and leave the comparatively less challenging problem of interpolating the pure delay $\tau$ as future work. More specifically, we aim to interpolate the time-aligned spectrum centered around the population-mean. Accordingly, the $i^\text{th}$ data point entering the composition of context or target set $\gC, \gT$ is given for a particular subject $s$ by
\begin{align*}
	\left(x_i,\ y^{(s)}_i\right) &= \left(x_i,\ m_i^{(s)}-\bar{m}_i\right),
\end{align*}
where
\begin{align}
	\bar{m}_i &= \frac{1}{S}\sum_{s=1}^S m^{(s)}_i, \label{eq:mean_envelope}
\end{align}
denotes the time-alined spectrum mean taken across the $S$ subjects of the train set and $m^{(s)}_i=m^{(s)}(x_i)$ denotes the value of time-alined spectrum specific to subject $s$ at location $x_i$.

\par Each task $(\gC, \gT)$ in the train/validate/test meta-set splits is composed using the HRIR filters from a single individual's set in the HUTUBS database~\cite{brinkmann2019a}. In particular, the context sets $\gC=\left\{(x_c,y_c)\right\}_{c=1}^C$ are of varying size and comprise from zero to a hundred data points sampled on the unit sphere according to an approximately-uniform-grid layout. In practice, one such approximately-uniform grid is prepared beforehand for each possible sample count. For each generated task $(\gC,\gT)$, one of these grids is randomly drawn, thereby selecting both the number of context data point samples and their relative locations on the unit sphere. Following this, a randomly-determined three-dimensional rotation of the grid is conducted to produce the final set of sampled coordinates on the unit sphere. Finaly, the HRTF set data points closest to the coordinates of the rotated grid are elected to form the context set $\gC$. The remaining data points of the HRTF set are used to form the target set $\gT$.

\par In order to augment the meta-train set, the uniform grid is replaced by an irregular grid with identical data point count half of the time during training. In particular, the coordinates of the irregular grid are in this case drawn independently following a uniform density across the surface of the sphere. Furthermore, the data points of the task $(\gC, \gT)$ are mirrored about the median plane half of the time. This augments the meta-train set with variants of the original subjects presenting permuted ears.%

\par Given that the simulated HRTF sets from the HUTUBS database comprise 1730 data points per subject, the approach described above provides a great number of interpolation tasks. In practice, each task $(\gC,\gT)$ is generated in real time within the train loop. This results in a meta-training set $\gM$ of considerable size from relatively few subjects. A summary of the HUTUBS subjects split among the meta-train, meta-validation and meta-test set is given in Table \ref{table:dataset_splits}. In this split, subjects 88 and 96 are discarded since they form duplicates of subjects 22 and 1 respectively~\cite{brinkmann2019hutubs}.
\begin{table}[t!]
	\caption{Split of subjects from the HUTUBS' simulated HRTF database~\cite{brinkmann2019a} \label{table:dataset_splits}}
	\centering
	\begin{tabular}{l | l | c}
		Set 			& Subjects			& Count  \\
		\hline
		Meta-train 		& \begin{tabular}{@{}c@{}}
							All but 1, 4, 18, 27, \\
							28, 53, 65, 67, 88, 96
						  \end{tabular} & 85\\
		Meta-validate	& 4, 28, 30, 53, 65 & 5\\
		Meta-test		& 1, 18, 27, 67 & 4
	\end{tabular}
\end{table}

\subsection{SConvCNP model}

\par The ConvCNP model was originally introduced with applications on planar data, such as images~\cite{gordon2019convolutional}. Accordingly, we adapt it to the spherical geometry of HRTF data and to the approximate symmetry between the left and right channels of the HRTF about the median plane. A detailed description of the resulting SConvCNP model is provided in this section.

\begin{figure}[t!]
\centering
		\includegraphics[trim=0 0 0 0, clip, width=3.5in]{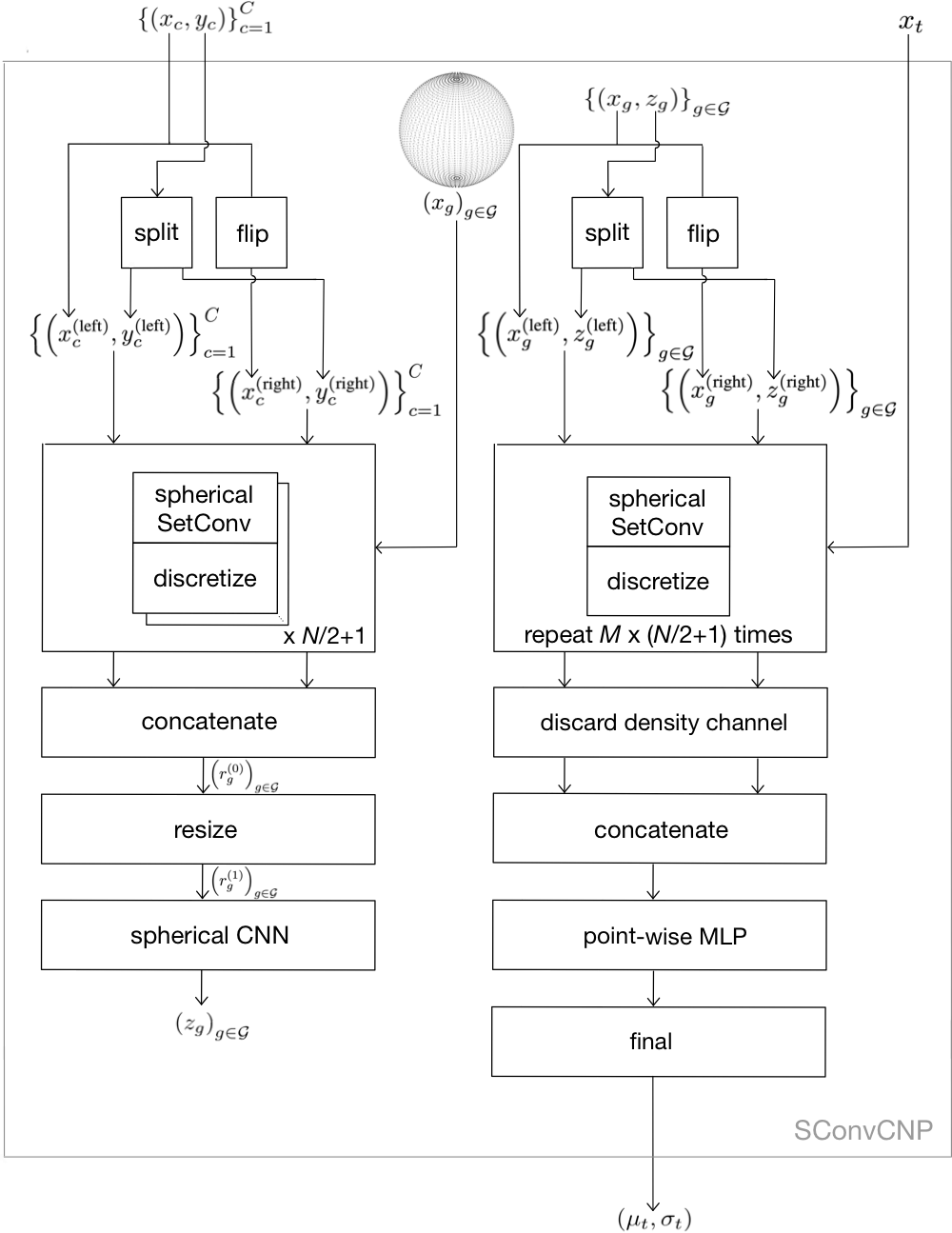}
\caption{Schematic block diagram of the SConvCNP model. Refer to Table \ref{table:tensor_dimensions} for tensor dimensions.\label{fig:sconvcnp}}
\end{figure}

\par Assuming a subject's morphology is perfectly symmetric about the median plane, the right HRTF channel would be perfectly recoverable from the left, thereby reducing the effective dimensionality of the HRTF feature space by a factor of two. In practice however, subjects are only approximately symmetric. Nevertheless, allowing observed feature values from one channel to inform the values in the opposite channel at the mirrored location should facilitate HRTF interpolation. The SConvCNP ensures this by mirroring the right channel of the data points about the median plane. As shown in Fig.~\ref{fig:sconvcnp}, the context set is decomposed (in the ``split'' block) into two channel-specific context sets at the input of the first discretized set convolution block. In particular, the coordinates perpendicular to the median plane are flipped (``flip'' block) in the right channel's context set. The set convolution processes each context set in sequence with shared parameters and the two resulting tensors are concatenated along the channel dimension downstream (``concatenate'' block). Fig.~\ref{fig:sconvcnp} also shows that the mirroring operation is executed a second time for the right channel, upstream of the second discretized set convolution. This recovers proper left-right filter channel pairings at the output.

\par A significant aspect of the interpolation task described in Sec. \ref{sec:hrtf_interpolation_task} lies in the spherical geometry of the time-aligned spectrum features to be interpolated. Specifically, each data point location takes value on the unit sphere. Accordingly, specialized set convolutions adapted to this spherical geometry are implemented in the SConvCNP model, as pictured in Fig. \ref{fig:sconvcnp}. In this work, we use a spherical Gaussian kernel~\cite{fasshauer2011positive}:
\begin{align}
	K(x_1, x_2) &= \operatorname{e}^{-2\beta (1-x_1\cdot x_2)}, \label{eq:spherical_gaussian}
\end{align}
where $x_1,\ x_2\in \left\{x \in \sR^3\ \left|\ \|x\|_2=1\right.\right\}$, $\cdot$ represents the dot product and the precision parameter $\beta\in(0,\infty)$ is learned. As pictured in Fig. \ref{fig:sconvcnp}, the first set convolution block carries out a dedicated spherical set convolution for each frequency bin, ensuring a specific precision parameter $\beta$ is learned at each frequency. In contrast, the second set convolution block performs a single discretized set convolution operation repeatedly with a single learned precision parameter $\beta$ shared across all channel-frequency pairs. Furthermore, the density channel at the output of the second set convolution is discarded (``discard density channel'' block).

\par To further accommodate the spherical geometry of HRTF data, we substitute planar convolutional layers in the CNN component with recently proposed spherical ones~\cite{cohen2018spherical, esteves2018learning}. Correspondingly, rotation equivariance is achieved in place of translation equivariance. We based our implementation on publicly-available code provided for spin-weighted spherical convolution \cite{esteves2020spin}\footnote{\url{https://github.com/google-research/google-research/tree/master/spin_spherical_cnns}}. In particular, we recover Esteves' simple zonal filter convolution \cite{esteves2018learning} as a special case discarding all spin directions but the null-valued one. In the resulting layer, the convolution operation is carried out in Spherical Harmonic (SH) space by matrix-multiplication of the input features with the layer's filter coefficients. In practice, the SH representation of the filter is interpolated directly from a few number of learnable SH coefficients. This provides localized zonal filters while simultaneously avoiding the cost of the forward SH transform for that part of the operation~\cite{esteves2018learning}.

\par In principle, the frequency dimension could be treated as an additional channel dimension. In this work, we propose to implement (single dimension) planar convolution in this dimension in order to promote the meta-learner's sample efficiency. This results in a three-dimensional hybrid planar-spherical convolution, with one axis for the frequency bins and two for the sound source direction. As reported in table \ref{table:tensor_dimensions}, the receptive fields has dimensions $G\times G\times (N/2+1)$ throughout all planar-spherical convolutional layers, where $G$ denotes the number of equiangular samples in each azimuth and elevation directions.

\begin{table}[t!]
	\caption{Dimension of tensors in the SConvCNP model\label{table:tensor_dimensions}}
	\centering
	\begin{tabular}{l | c | c | c}
		Tensor 			& Type & Shape & Channels\\
		\hline
		$y_c$	& $\mathds{C}$ & $\left(N/2 +1\right)$& $2$\\
		$y_c^{(\text{left})}$	& $\mathds{C}$ & $\left(N/2 +1\right)$&$ 1$\\
		$y_c^{(\text{right})}$	& $\mathds{C}$ & $\left(N/2 +1\right)$&$ 1$\\
		$\left(r_g^{(0)}\right)_{g\in\gG}$		& $\mathds{C}$ & $G \times G\times \left(N/2 +1\right) $& $4$\\
		$\left(r_g^{(\ell)}\right)_{g\in\gG},\ \ell>0$		& $\mathds{R}$ & $G \times G\times \left(N/2 +1\right) $&$ M$\\
		$\left(z_g\right)_{g\in\gG}$		& $\mathds{R}$ & $G\times G\times\left(N/2 +1\right) $&$ M$\\
	$\left(z_g^{(\text{left})}\right)_{g\in\gG}$ & $\mathds{R}$ & $G \times G\times \left(N/2 +1\right)  $&$M/2$\\
	$\left(z_g^{(\text{right})}\right)_{g\in\gG}$ & $\mathds{R}$ & $G \times G\times \left(N/2 +1\right)  $&$M/2$\\
		$q_t^{(\ell)}$		& $\mathds{R}$ & $\left(N/2 +1\right) $&$ M$\\
		$\mu_t$ 	&$\mathds{C}$ & $\left(N/2 +1\right) $&$ 2$\\
		$\sigma_t$ 	& $\mathds{C}$ & $\left(N/2 +1\right) $&$ 2$ %
	\end{tabular}
\end{table}

\par
\begin{figure}[t!]
\centering
	\begin{tabular}{c c}
		\includegraphics[width=0.7in, clip]{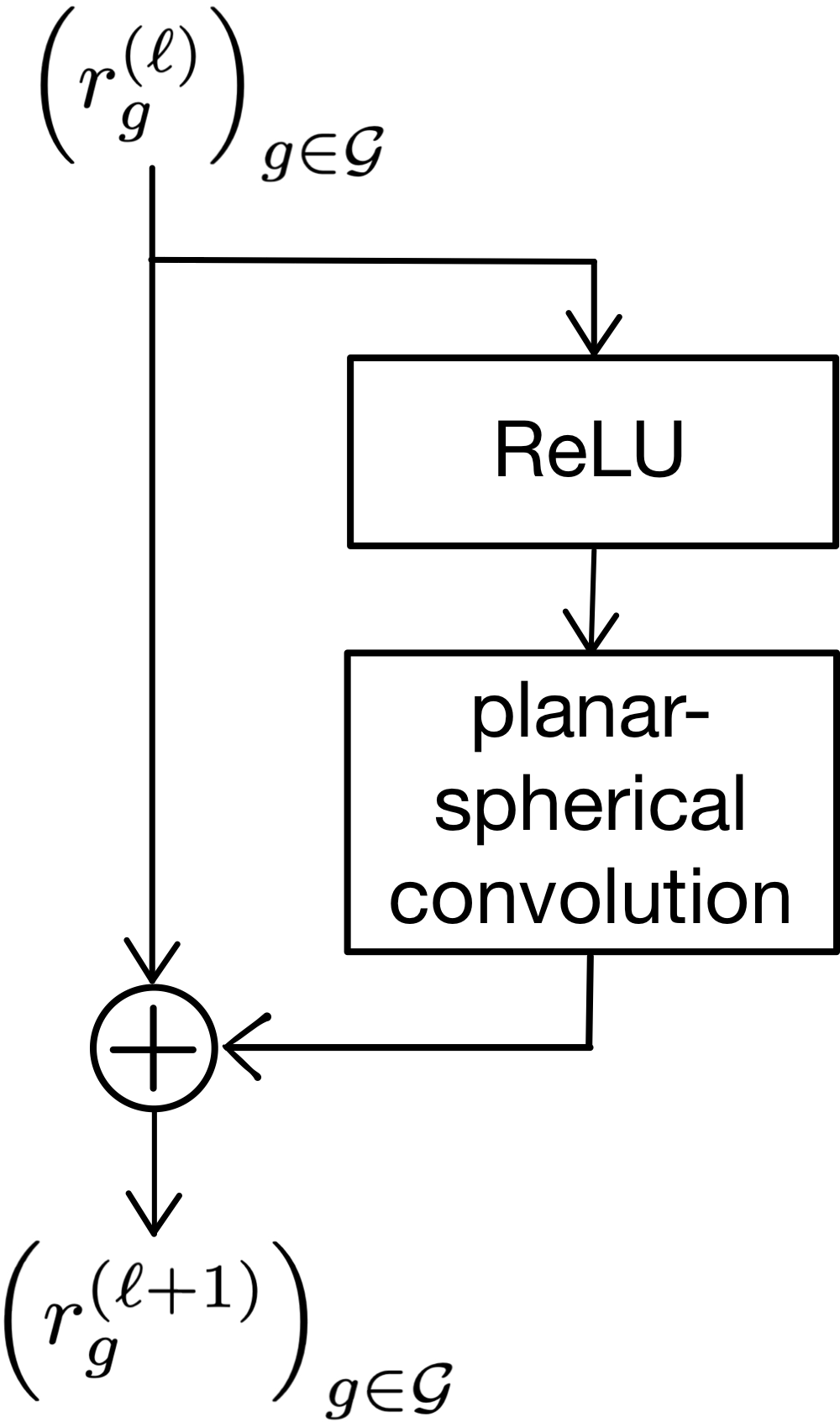}&\qquad \qquad \qquad
		\includegraphics[width=0.7in, clip]{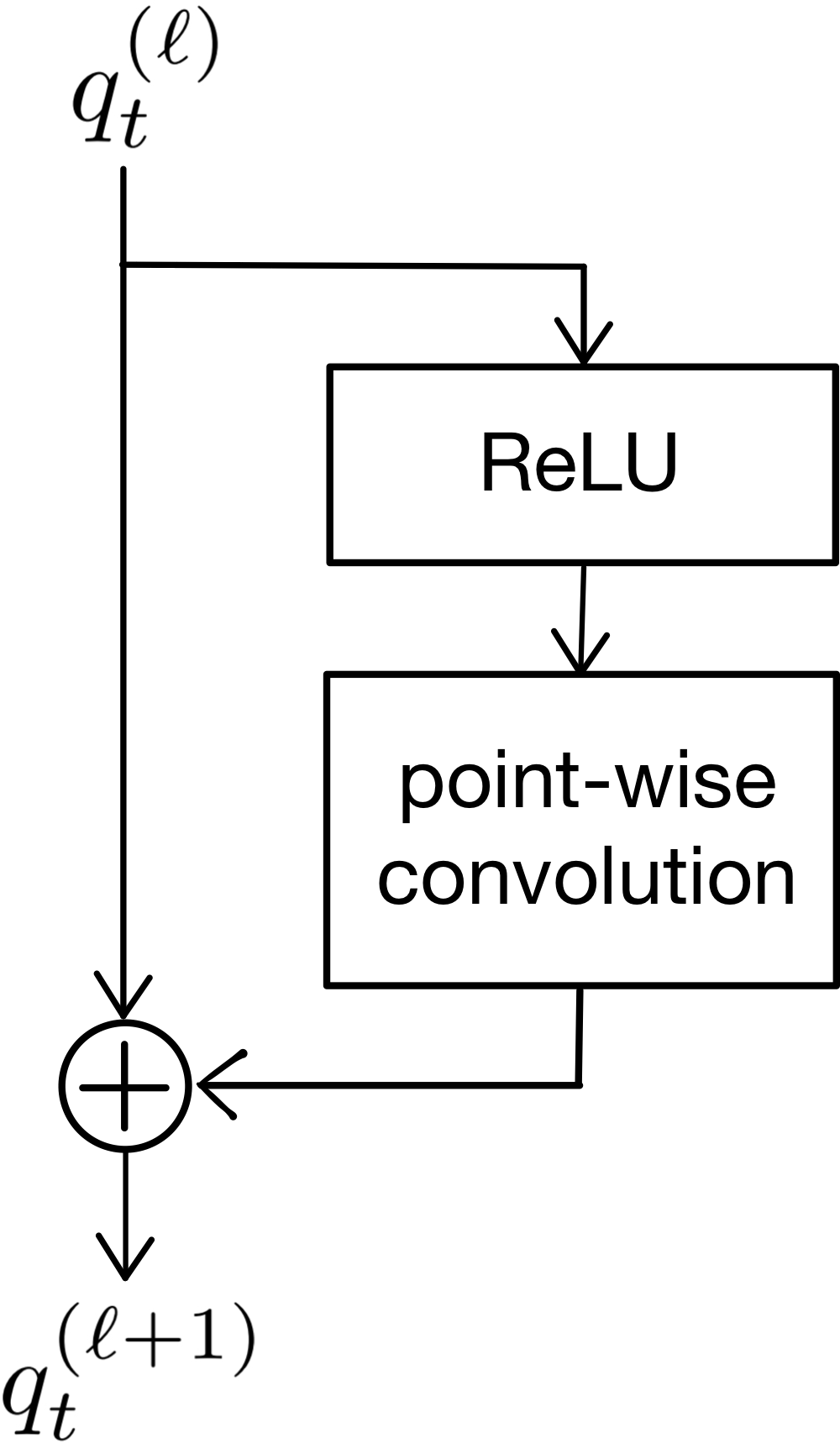}
	\end{tabular}
\caption{Single-layer pre-activation residual blocks used to compose the spherical CNN (left) and MLP (right) components of the SConvCNP model of Fig.~\ref{fig:sconvcnp}. Refer to Table \ref{table:tensor_dimensions} for dimension of tensors.\label{fig:residual_blocks}}
\end{figure}

 In classical fashion, the spherical CNN component of the model is composed of residual blocks arranged in a sequence. As pictured in Fig.~\ref{fig:residual_blocks}, each block follows a single-layer pre-activation architecture~\cite{he2016identity}. As common in residual architectures, a ``resize'' layer is positioned at the input of the spherical CNN. Firstly, this block converts the complex-valued input tensor into an equivalent float-valued tensor, by concatenating real and imaginary parts along the channel dimension. Moreover, it scales the number of channels to the specified count M used in the residual blocks of the spherical CNN.

\par The point-wise MLP component of the SConvCNP model is also composed of single-layer pre-activation residual blocks as represented in Fig.~\ref{fig:residual_blocks}. Each block is implemented using a point-wise convolution layer for sharing parameters across frequency bins. %
The model comprises a final layer that resizes and splits the channel dimension to provide a complex-valued predictive mean tensor $\mu_t$ and an unconstrained complex-valued standard deviation tensor $\sigma_t^\prime$. Furthoremore, this layer provides the predictive standard deviation tensor $\sigma_t$ using a risen $\operatorname{softplus}$ non-linearity forcing the real and imaginary parts of the unconstrained standard deviation coefficients to positive values~\cite{le2018empirical,gordon2019convolutional}:
\begin{align*}
	\sigma_t &= \operatorname{risen\_softplus}\left(\operatorname{Re}\left(\sigma_t^\prime\right)\right) + \text{i}\operatorname{risen\_softplus}\left(\operatorname{Im}\left(\sigma_t^\prime\right)\right),
\end{align*}
where
\begin{align*}
	\operatorname{risen\_softplus}\left(\nu\right) &= 	\sigma_\text{floor} + \left(1-\sigma_\text{floor}\right)\log\left(1+\operatorname{e}^\nu\right),
\end{align*}
and $\sigma_\text{floor}\in (0,\infty)$ is small. This yields the following conditional probability density estimate for the target features $y_t$:
\begin{align*}
	p\left(y_t\ \left|\ x_t,\ \left\{(x_c,y_c) \right\}_{c=1}^C\right.\right) \approx \hdots \qquad \qquad \qquad \qquad \qquad\qquad \\
	\gN\left(y_t^{\operatorname{Re}};\ 
			\mu_t^{\operatorname{Re}},\ 
			\operatorname{diag}\left(\sigma_t^{\operatorname{Re}}\right)^2
		\right)
	\gN\left(y_t^{\operatorname{Re}};\ 
			\mu_t^{\operatorname{Im}},\ 
			\operatorname{diag}\left(\sigma_t^{\operatorname{Im}}\right)^2
		\right),
\end{align*}
where $\gN$ denotes the multivariate normal distribution, $y_t^{\operatorname{Re}}=\operatorname{flatten}\left(\operatorname{Re}\left(y_t\right)\right)$, $\mu_t^{\operatorname{Re}}=\operatorname{flatten}\left(\operatorname{Re}\left(\mu_t\right)\right)$, 
$\sigma_t^{\operatorname{Re}}=\operatorname{flatten}\left(\operatorname{Re}\left(\sigma_t\right)\right)$, $y_t^{\operatorname{Im}}=\operatorname{flatten}\left(\operatorname{Im}\left(y_t\right)\right)$, $\mu_t^{\operatorname{Im}}=\operatorname{flatten}\left(\operatorname{Im}\left(\mu_t\right)\right)$, $
\sigma_t^{\operatorname{Im}} = \operatorname{flatten}\left(\operatorname{Im}\left(\sigma_t\right)\right)$, and $\operatorname{flatten}$ reshapes the tensor provided as argument into a vector.

\subsection{Interpolation Accuracy}

\par In this work, we compare the performance of the \mbox{SConvCNP} model to Gaussian process regressor, thin-plate spherical spline and barycentric interpolation baselines. Interpolation is carried out for all three methods on the \mbox{SConvCNP} model's input features. Similarly to \cite{carlile2002performance}, the thin-plate spherical spline method is implemented following Whaba~\cite{wahba1981spline}, using second-order splines and without smoothing. Gaussian process regression is conducted on a per-frequency basis similarly to Luo et al.~\cite{luo2013kernel}. At each frequency, we define a covariance function for the real part and one for the imaginary part of the bin using the spherical Gaussian kernel from Equation (\ref{eq:spherical_gaussian}). The observational noise of the model is fixed with a value of 1e-4. The remaining meta-parameters, in particular the precision parameters from the spherical Gaussian kernels, are fitted on 340 tasks from the meta-train set under the log marginal likelihood objective~\cite{rasmussen2006gaussian}. Meta-parameter values are maintained fixed upon evaluation on the meta-test set. In classical fashion, the barycentric interpolation baseline provides each interpolated feature $\hat{y}$ as a convex combination of the values $\left\{y_i\right\}_{i=1}^3$ found at the observed data points defining the smallest spherical triangle enclosing the target point location $x$, i.e.:
\begin{align*}
	\hat{y} &= \sum_{i=1}^3 b_i y_i,
\end{align*}
where $b_i$ denotes the barycentric coordinate of the target location $x$ associated with the $i^\text{th}$ vertex of said spherical triangle. In classical fashion, the barycentric coordinates $b_i$ are computed as ratios of spherical triangle areas, each computed as the sum of the spherical angles.

\par We also compare the SConvCNP model's HRTF magnitude interpolation performance specifically, to that of a publicly-available implementation\footnote{\url{https://github.com/AudioGroupCologne/SUpDEq}} of the natural-neighbors interpolation method~\cite{porschmann2020comparison,arend2023magnitude}. In particular, we apply this implementation directly on the HRTF spectrum after downsampling to 33.075 kHz but without any time-alignment pre-processing. More specifically, we run the implementation provided for the NAT-PH variant, which carries out interpolation on the magnitude and phase of the HRTF as described in \cite{porschmann2020comparison}.

\par Candidate methods are compared using common metrics computed on a per-feature basis, including the relative error (LRE)
\begin{align}
	\text{LRE}\left(m_{f,e}, \hat{m}_{f,e}\right) &= 20\log_{10} \left|\frac{\hat{m}_{f,e} - m_{f,e}}{m_{f,e}}\right|,\label{eq:relative_error}
\end{align}
and the log-magnitude distance (LMD)
\begin{align}
	\text{LMD}\left(m_{f,e}, \hat{m}_{f,e}\right) &= \left|20 \log_{10} \left|\frac{\hat{m}_{f,e}}{m_{f,e}}\right|\right|,\label{eq:log_magnitude_distance}
\end{align}
where in a slight departure of notation, $\hat{m}$ and $m$ denote here the predicted point-wise time-aligned HRTF spectrum value and the ground truth value respectively, $f$ indexes over the frequency bin, and $e$ indexes over the left and right ears. For completeness, we also report log-spectral distortion (LSD) which is given in prior work as follows for a whole binaural filter~\cite{ito2022head}: 
\begin{align}
	\text{LSD}\left(m, \hat{m}\right) &= \frac{1}{2}\sum_{e=1}^{2}
\sqrt{\frac{1}{\left(N/2 + 1\right)}\sum_{f=1}^{N/2 + 1}\left(20 \log_{10} \left|\frac{\hat{m}_{f,e}}{m_{f,e}}\right|\right)^2}. \label{eq:log_spectral_distortion}
\end{align}

\subsection{Uncertainty Calibration}
\par Several methods have been proposed for assessing a regressor's ability to gauge the uncertainty it provides alongside its point-wise predictions~\cite{gneiting2007probabilistic, kuleshov2018accurate, levi2022evaluating}. In particular, Levi et al. introduce a particular definition of uncertainty calibration according to which the model is calibrated if, in expectation over the data-generating distribution, the predicted variance it provides matches the squared error it commits upon carrying out the point-wise prediction~\cite{levi2022evaluating}. In principle, this condition must hold across all possible values for the predicted variance.

\par In practice, an approximate but tractable verification of this condition can be conducted for a limited number of variance values using a data set of finite size~\cite{levi2022evaluating}. In such an approach, the resulting set of predicted variance and squared error pairs are divided into equally-sized groups forming non-overlapping contiguous interval divisions of the predicted variance axis. The expectation over the data-generating distribution is approximated within each group as the sample mean of squared error values in the group. The resulting mean squared error (MSE) values obtained for all groups are plotted as a function of the groups' respective mean predicted variance values (MPV). This allows for assessing the degree of miss-calibration. In particular, overconfident models produce a MPV versus MSE curve exceeding the identity line. Under-confident ones produce a curve lying under it. Miss-calibration can be summarized by a single-scalar mean-aggregate of the calibration error~\cite{levi2022evaluating}. In this work we propose to use the following mean calibration distance (MCD) metric:
\begin{align}
\frac{1}{D}\sum_{i=1}^D \left|10\log_{10}\frac{\text{MSE}_i}{\text{MPV}_i}\right|,\label{eq:mcd}
\end{align}
where $D$ denotes the number of divisions of the predicted variance axis.

%% file: sections/results.tex
\par This section summarizes the meta-test set performance of a selected SConvCNP model configuration, which, among other candidates, achieved, after early stopping, near-best meta-validation set performance in both mean relative error level and mean calibration distance metrics according to (\ref{eq:relative_error}) and (\ref{eq:mcd}) respectively. All candidate configurations were trained with a batch size of 8. Both meta-test and meta-validation sets comprised 340 tasks. The selected configuration's spherical CNN and point-wise MLP components both comprise five residual blocks with $M=128$ channels each. The spherical convolution is implemented using a 64$\times$64 equiangular grid ($G=64$). Each planar-spherical filter is composed of 7 taps of SH representations interpolated from 16 learnable SH coefficients each~\cite{esteves2018learning}. The standard deviation floor $\sigma_\text{floor}$ value is 1e-4 in the selected model.

\begin{figure}[t!]
\centering
\begin{tabular}{@{}c@{}}
	\begin{tabular}[c]{@{}c@{}c@{}}
		\begin{tabular}[c]{@{}c@{}}
			\includegraphics[trim=0 40 0 40, clip, width=2.6in]{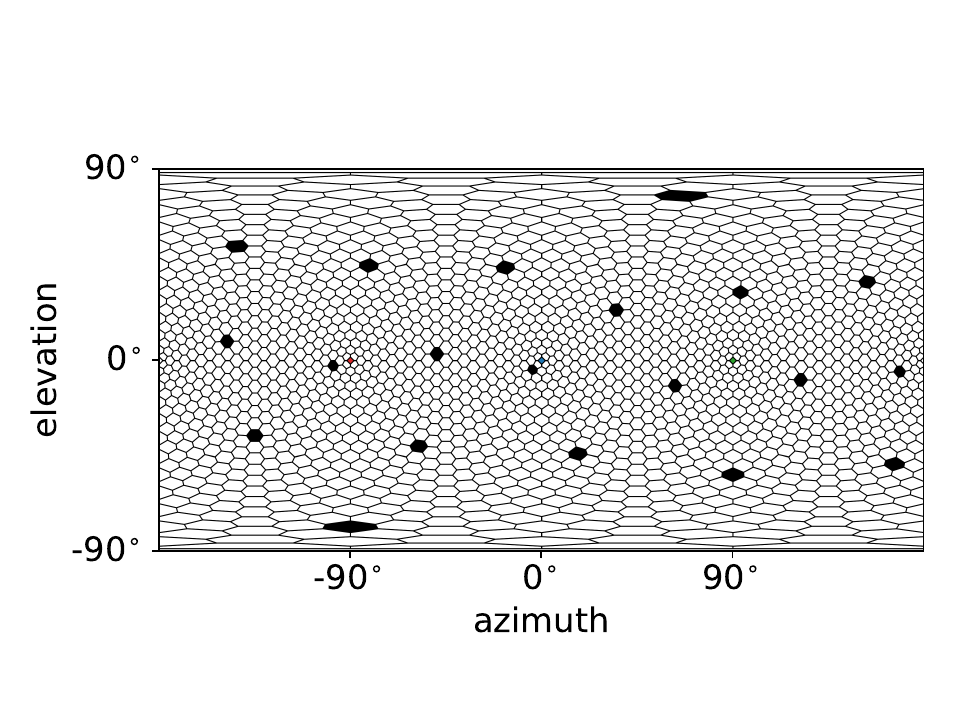}
		\end{tabular}
		&
		\begin{tabular}[c]{@{}c@{}}
			\includegraphics[trim=60 50 47 45, clip, width=0.9in]{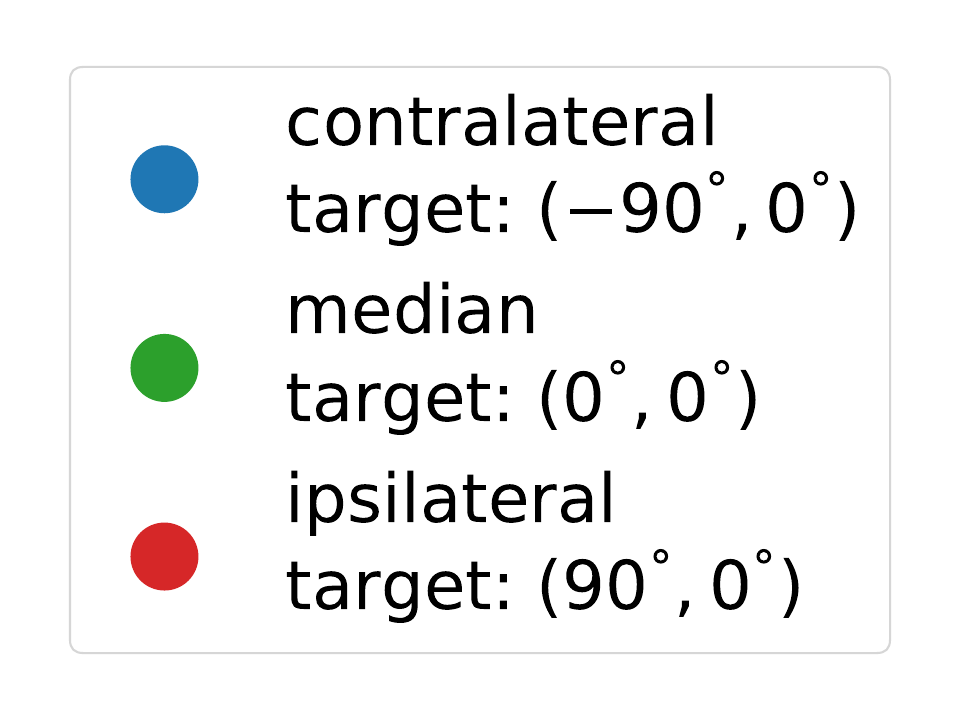}
		\end{tabular}
	\end{tabular}
	\\
	\begin{tabular}[c]{@{}c@{}c@{}}
		\includegraphics[trim=15 15 5 15, clip, width=1.77in]{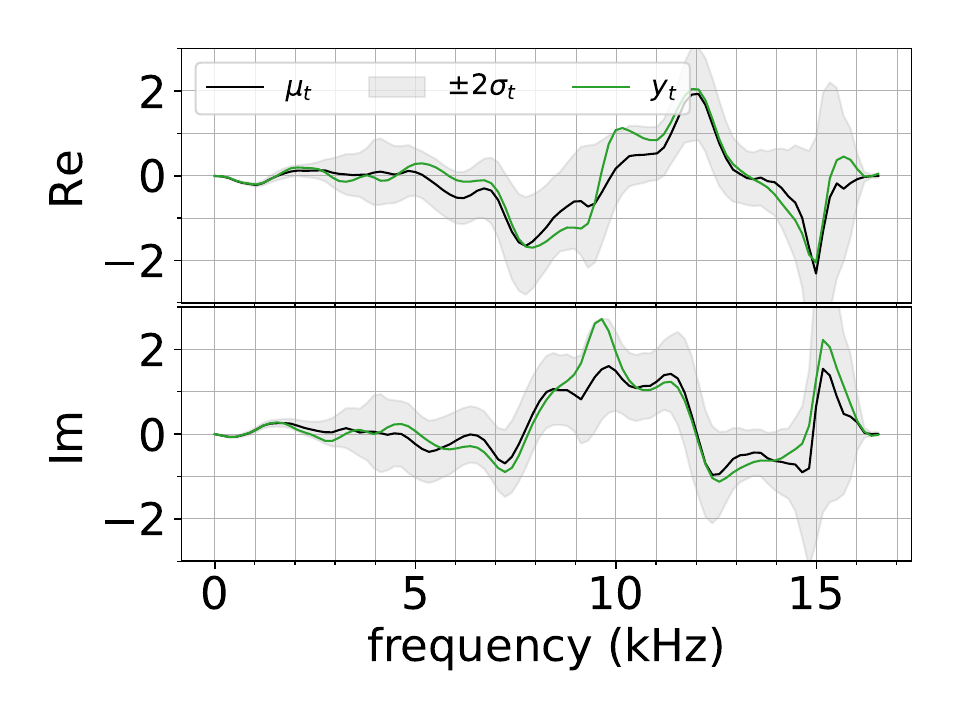}
		&
		\includegraphics[trim=15 15 5 15, clip, width=1.77in]{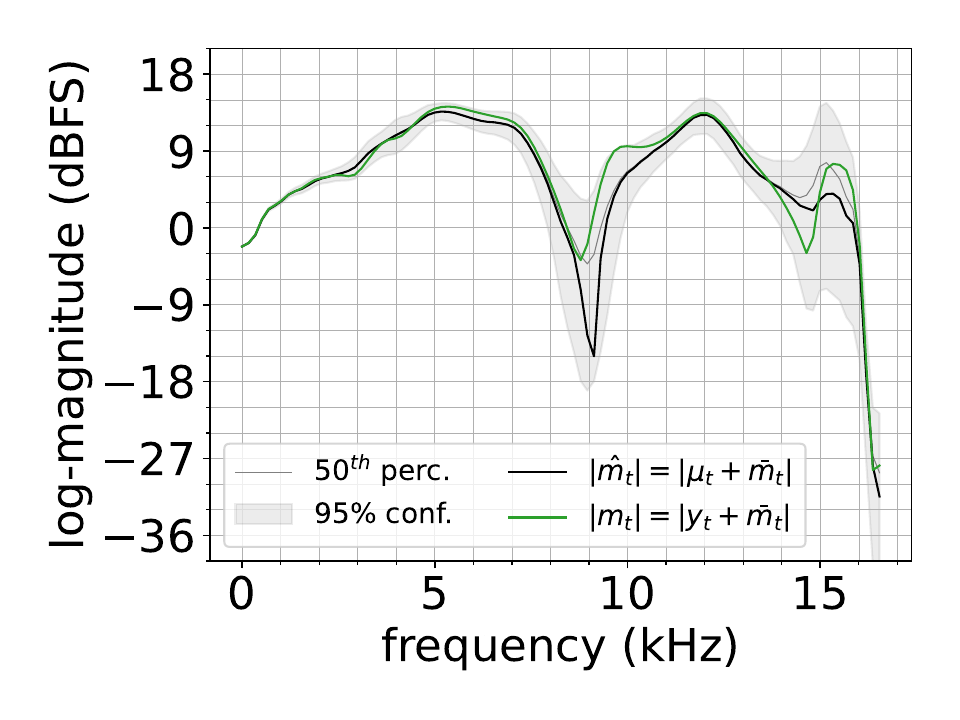}
		\\
		\includegraphics[trim=15 15 5 15, width=1.77in]{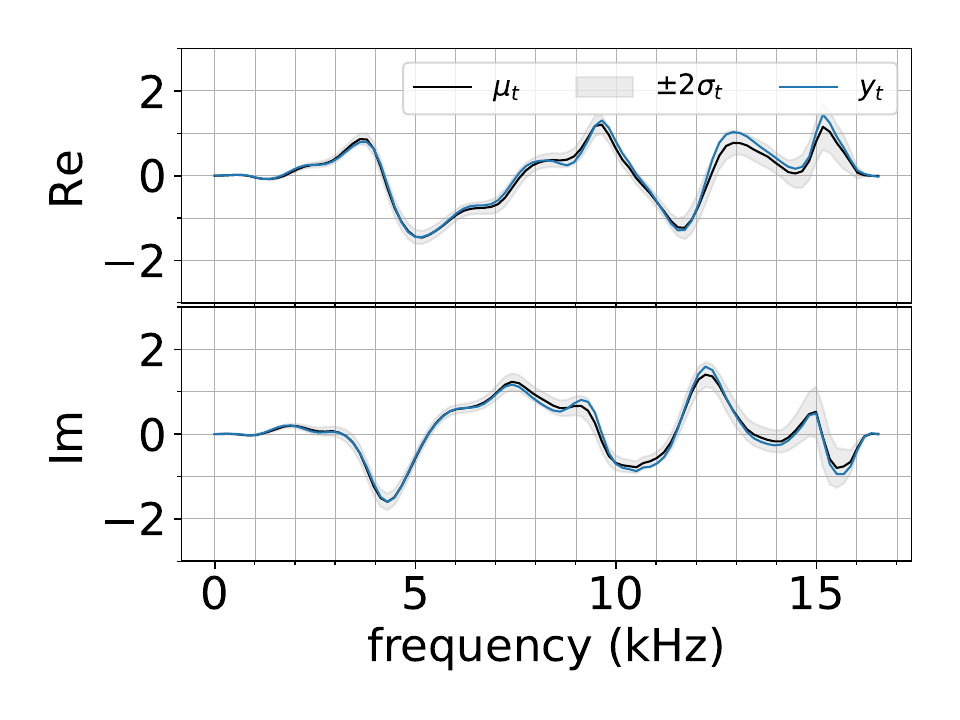}
		&
		\includegraphics[trim=15 15 5 15, clip, width=1.77in]{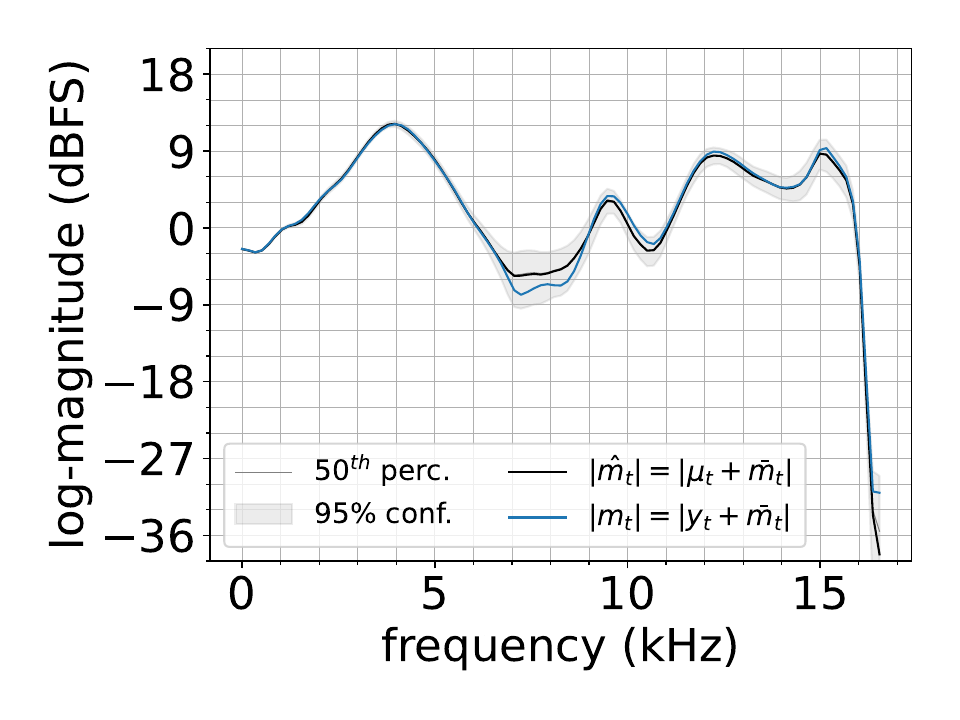}
		\\
		\includegraphics[trim=15 15 5 15, width=1.77in]{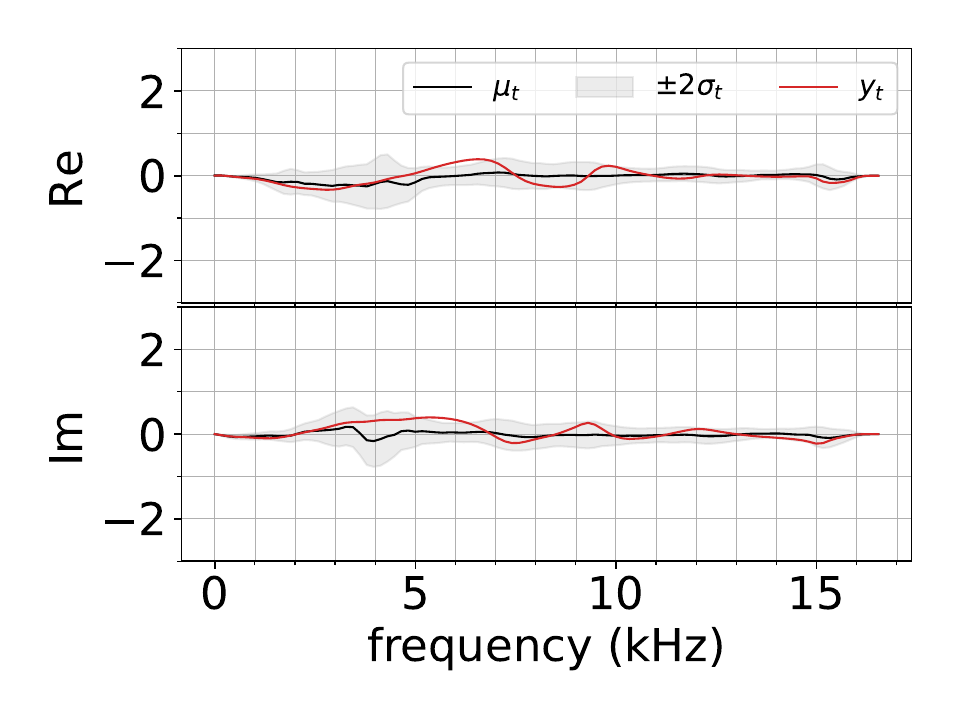}
		&
		\includegraphics[trim=15 15 5 15, clip, width=1.77in]{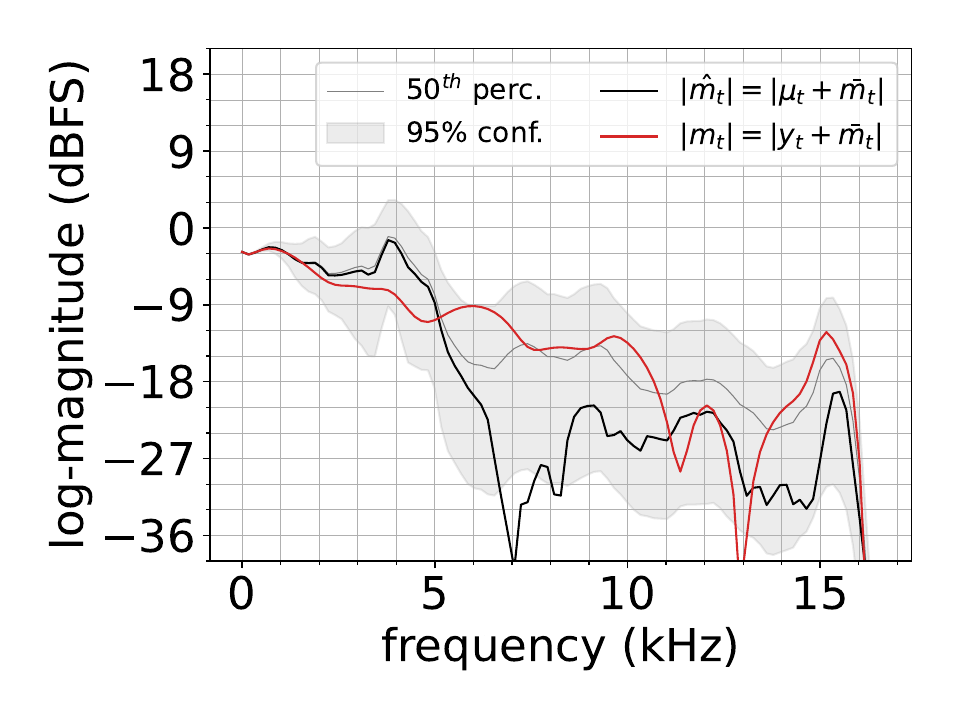}
	\end{tabular}
\end{tabular}
\caption{Time-aligned HRTF spectrum interpolation task example presenting twenty context data points sampled from subject 1 of the HUTUBS database. First row: diagram marking the locations of the context data points (black) as well as three target locations (colored). Left plots: ground truth residual time-aligned HRTF spectrum (colored) and corresponding predictive distribution (black, grey) provided by the SConvCNP for the left channel at the target locations. Right plots: corresponding log-magnitude HRTF spectrum.}%
\label{fig:log_magnitude_spectrum_example}
\end{figure}

\begin{figure}[t!]
\centering
\begin{tabular}{@{}c@{}}
	\begin{tabular}[c]{@{}c@{}c@{}}
		\includegraphics[trim=15 15 5 15, width=1.77in]{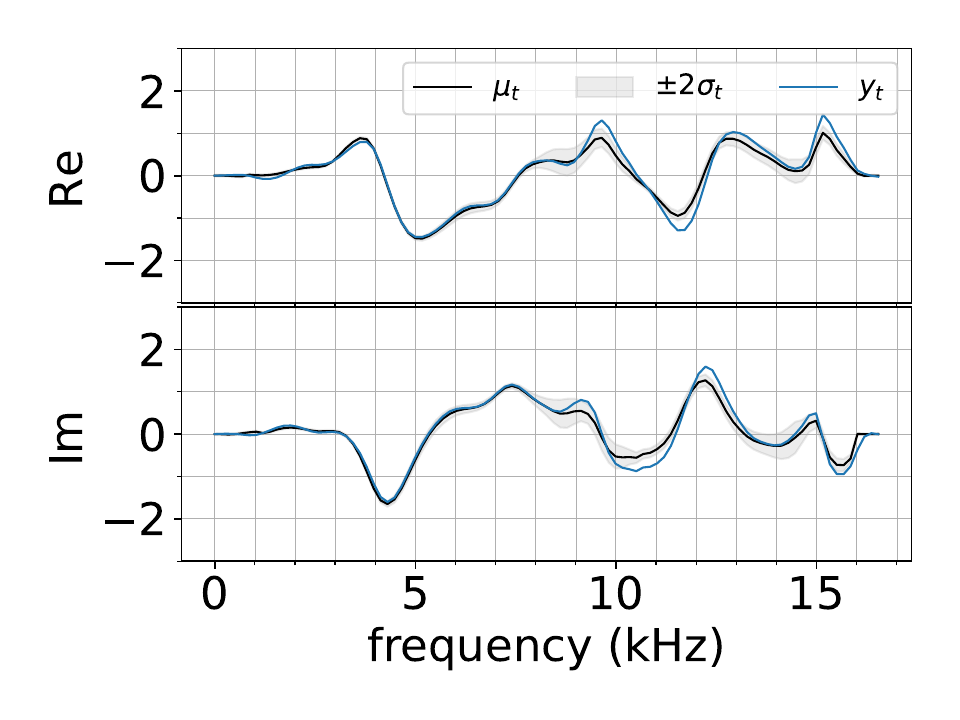}
		&
		\includegraphics[trim=15 15 5 15, clip, width=1.77in]{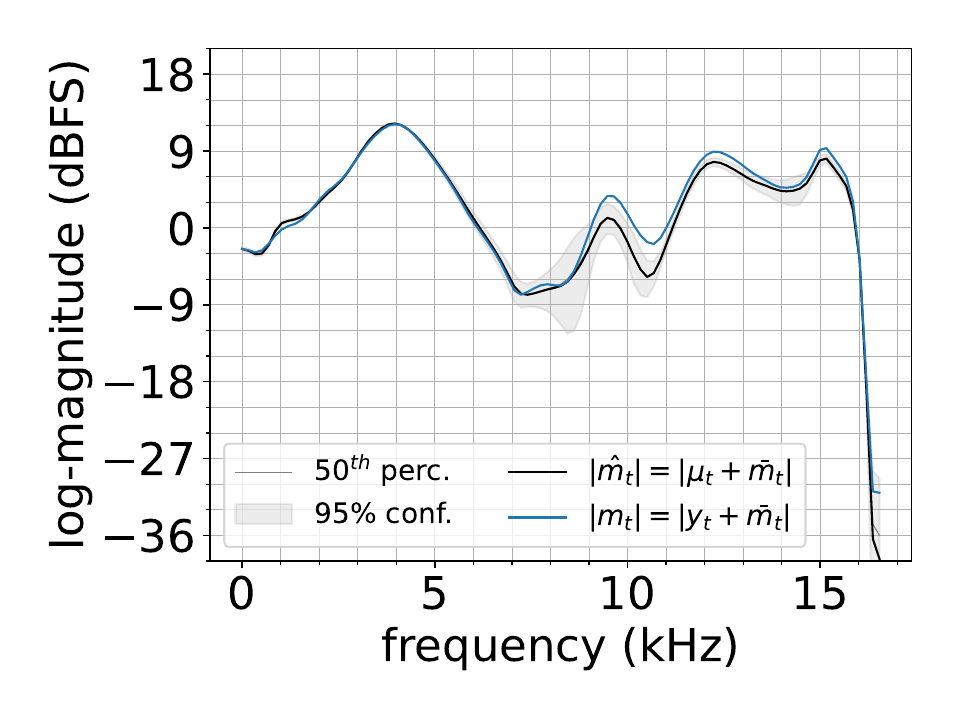}
		\\
		\includegraphics[trim=15 15 5 15, width=1.77in]{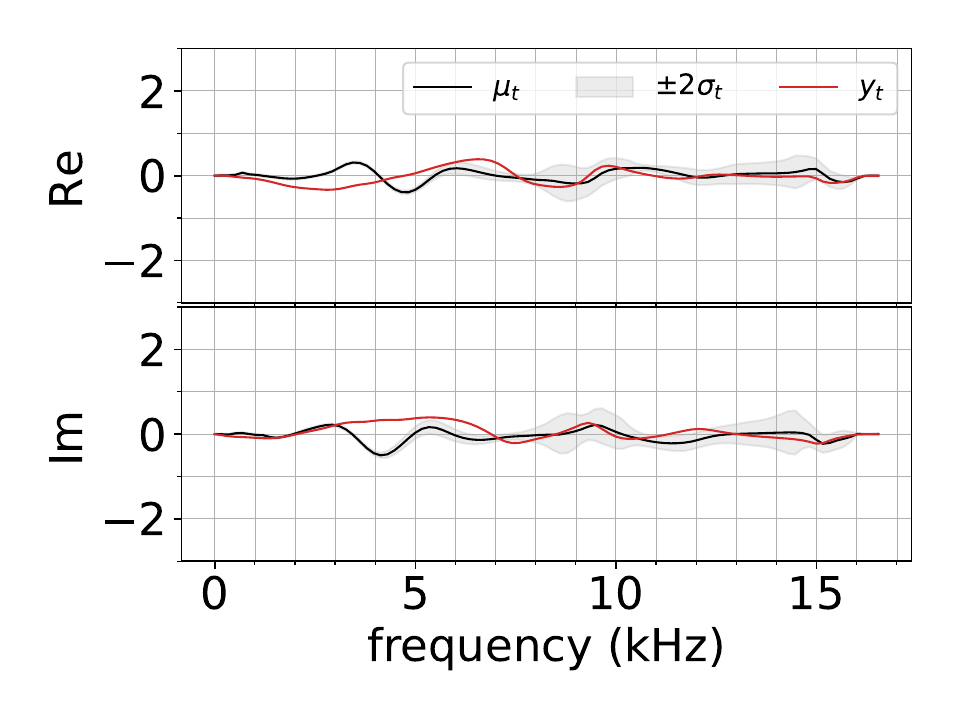}
		&
		\includegraphics[trim=15 15 5 15, clip, width=1.77in]{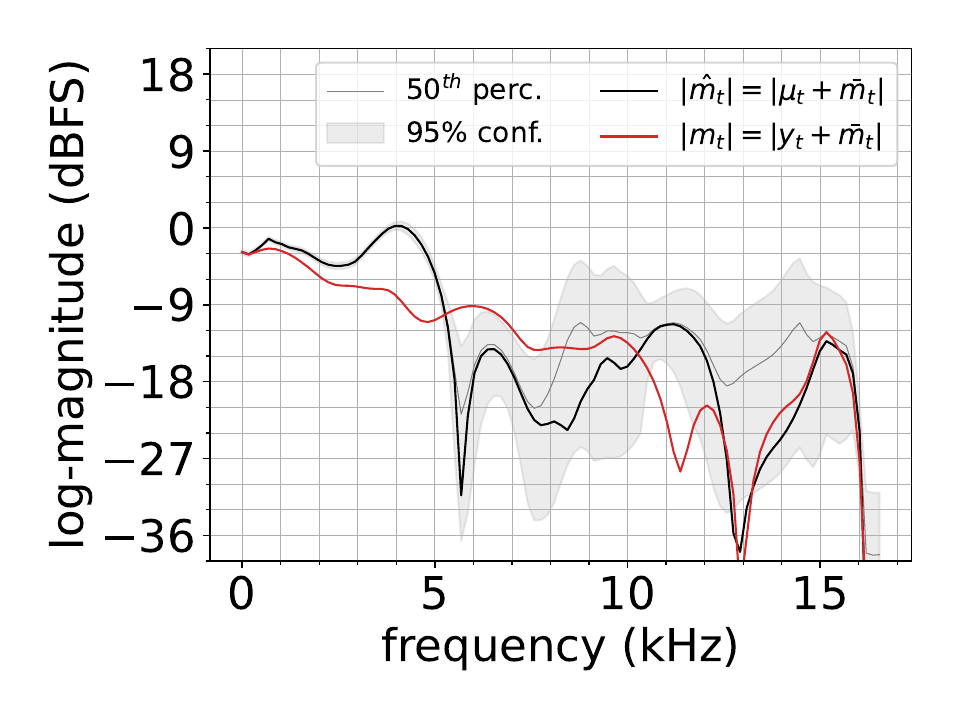}
	\end{tabular}
\end{tabular}
\caption{Left ear channel Gaussian Process solution at the frontal (top) and contralateral (bottom) directions for the interpolation task of Fig.~\ref{fig:log_magnitude_spectrum_example}.\label{fig:log_magnitude_spectrum_example_4_gp}}
\end{figure}

\par Fig. \ref{fig:log_magnitude_spectrum_example} provides an example of HRTF interpolation task using the SConvCNP candidate. In particular, this example is given for the FABIAN head and torso simulator (subject 1 of the HUTUBS dataset) and a specific draw of 20 context point locations represented in the top diagram of the figure (black markers). The diagram further marks the location of three target locations each with a distinct color code. The prediction provided by the SConvCNP model at each target location is reported in a corresponding row of the figure. The left plot of each row compares the predictive distribution to the ground truth. The right plot represents the associated log-magnitude spectra, namely point-wise estimate $\hat{m}=\mu_t+\bar{m}_t$ and ground truth $m=y_t+\bar{m}_t$ where $\bar{m}_t$ denotes the population mean as defined in (\ref{eq:mean_envelope}). The 50$^\text{th}$ percentile (median) and the 95\% confidence intervals appearing in the log-magnitude plots are simulated estimates, computed from a population of samples randomly drawn according to the predictive distribution provided by the SConvCNP model.

\par As pictured, the model's predictive distribution lies in good agreement with ground truth features $y_t$. In particular, the ground truth generally falls within the 95\% confidence interval ($\pm 2\sigma_t$ range, grey region) around the predictive mean in all three target location cases. Moreover, the predictive mean $\mu_t$ of the model (full black line) shows significant correlation with the ground truth $y_t$ in both the ipsilateral direction case (green marker)  and frontal direction case (blue marker). Furthermore, the model's prediction is more uncertain when the target point (ipsilateral direction, green marker) lies further away from context data points than in close vicinity (frontal direction, blue marker). This suggests the model's predictive distribution effectively captures model uncertainty.%

\par In contrast, the predictive mean is much less correlated with the ground truth in the contralateral direction case (red marker) despite the target direction being close to a context data point as indicated by the diagram at the top of Fig. \ref{fig:log_magnitude_spectrum_example}. In particular, the predictive mean is practically agnostic above the 5-kHz mark, with a near-zero value throughout, and the standard deviation extends significantly outwards from the abscissa to capture variations in ground truth value. This is not unexpected as interpolation is a harder problem in the contra-lateral region, where the HRTF is spatially more intricate such that correlations between data points would occur within small distances only. Given this, the magnitude spectrum estimate $\left|\hat{m}_t\right|=\left|\mu_t+\bar{m}_t\right|$ significantly undershoots the ground truth (right plot), while the transformed distribution's median (50\% percentile) better predicts the power spectrum of the filter in this case (red curve).

\par Fig.~\ref{fig:log_magnitude_spectrum_example_4_gp} depicts the solution provided by the gaussian process regressor baseline for the frontal and contralateral target directions of Fig.~\ref{fig:log_magnitude_spectrum_example}'s task. Contrary to SConvCNP, the Gaussian process' predictive distribution does not capture the variability of the ground truth features along the frequency axis in either case (blue or red curves). Crucially, the uncertainty estimates in each case are of similar value at any given frequency. This is expected since the Gaussian process' predictive uncertainty solution is a function of distance to context data points~\cite{rasmussen2006gaussian}, which is similar for both target locations given in the example. In this baseline specifically, the modeled degree of correlation between feature values at distinct locations on the unit sphere is tuned by the precision meta-parameter of the covariance function, and is hence equal in the contralateral and ipsilateral regions. This contrasts starkly with the SConvCNP model's ability to take account of observed feature values to provide well-behaved uncertainty estimates in regions exhibiting different degrees of spatial variability.

\par
\begin{figure}[!htbp]
\centering
\begin{tabular}{@{}c@{}}
58$^\text{th}$ frequency bin: $\sim$10 kHz\\
\begin{tabular}{@{}c@{}}
	\begin{tabular}[c]{@{}c@{}c@{}c@{}}
		\begin{tabular}[c]{@{}c@{}}
			$\qquad \left|m_t\right|$\\
			\includegraphics[trim=18 30 0 70, clip, width=1.6in]{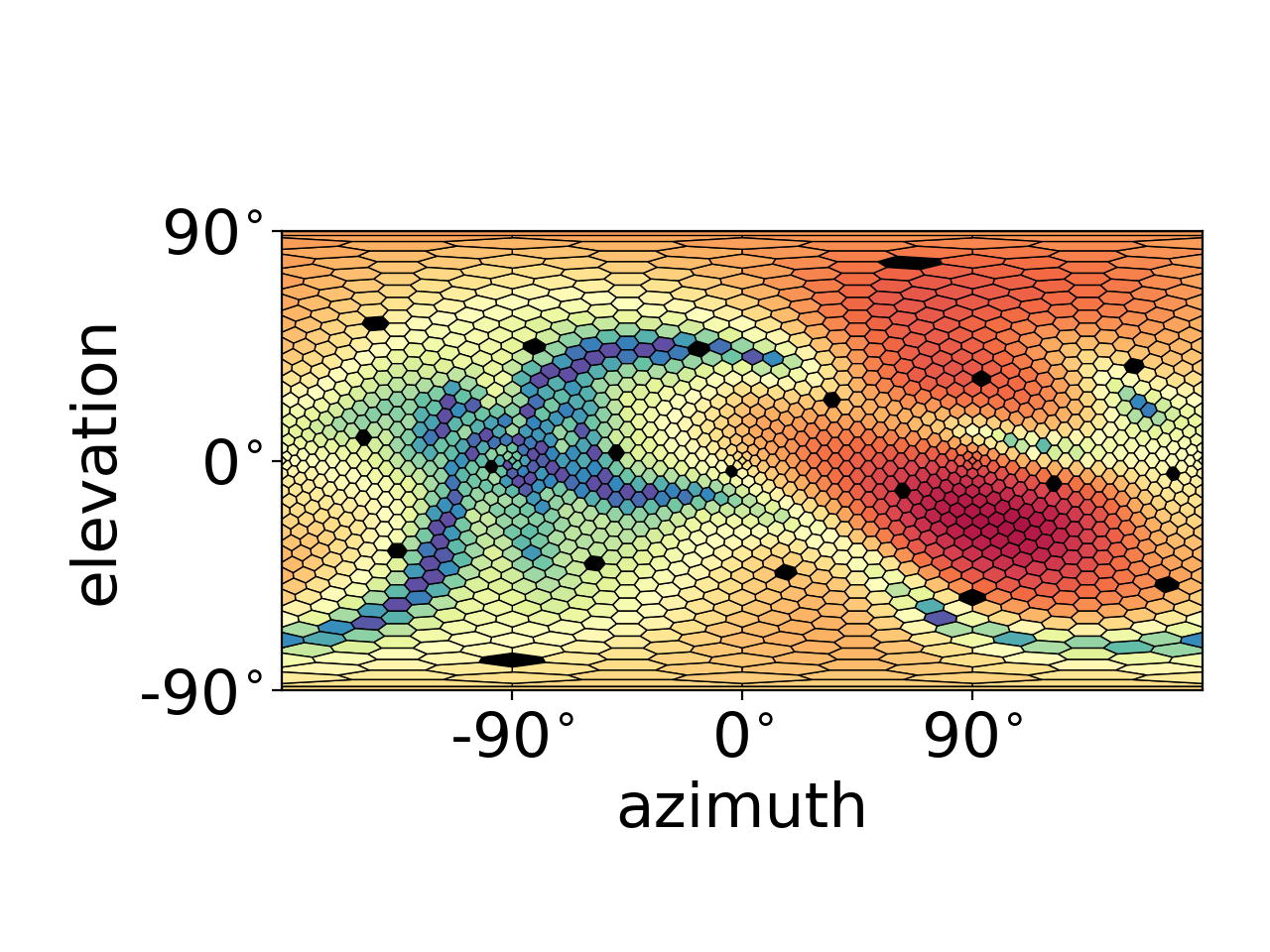}
		\end{tabular}
		&
		\begin{tabular}[c]{@{}c@{}}
			$\qquad \left|\hat{m}_t\right|$\\
			\includegraphics[trim=18 30 0 70, clip, width=1.6in]{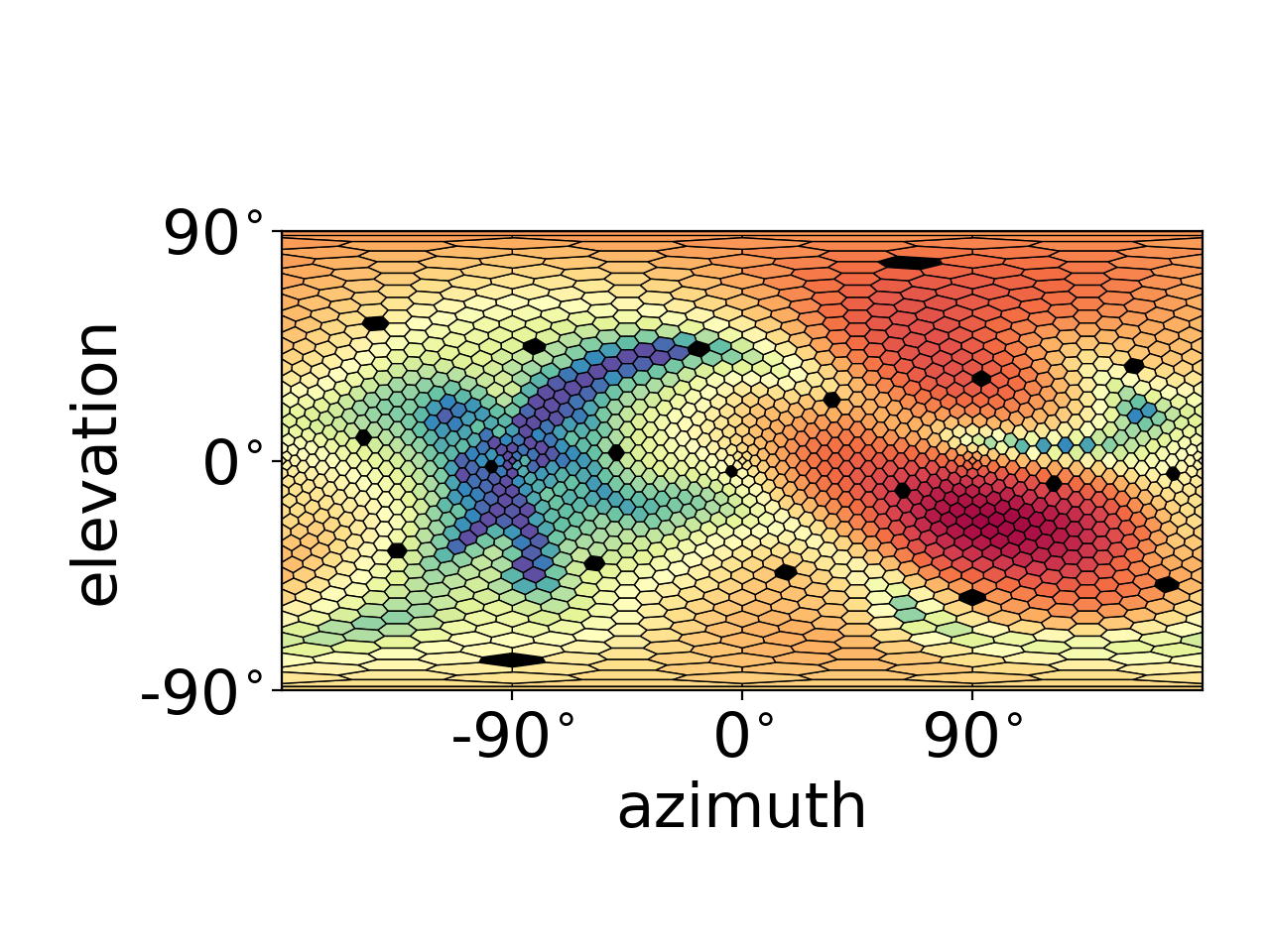}
		\end{tabular}
		&
		\begin{tabular}[c]{@{}c@{}}
			\includegraphics[trim=320 15 0 15, clip, width=0.4in]{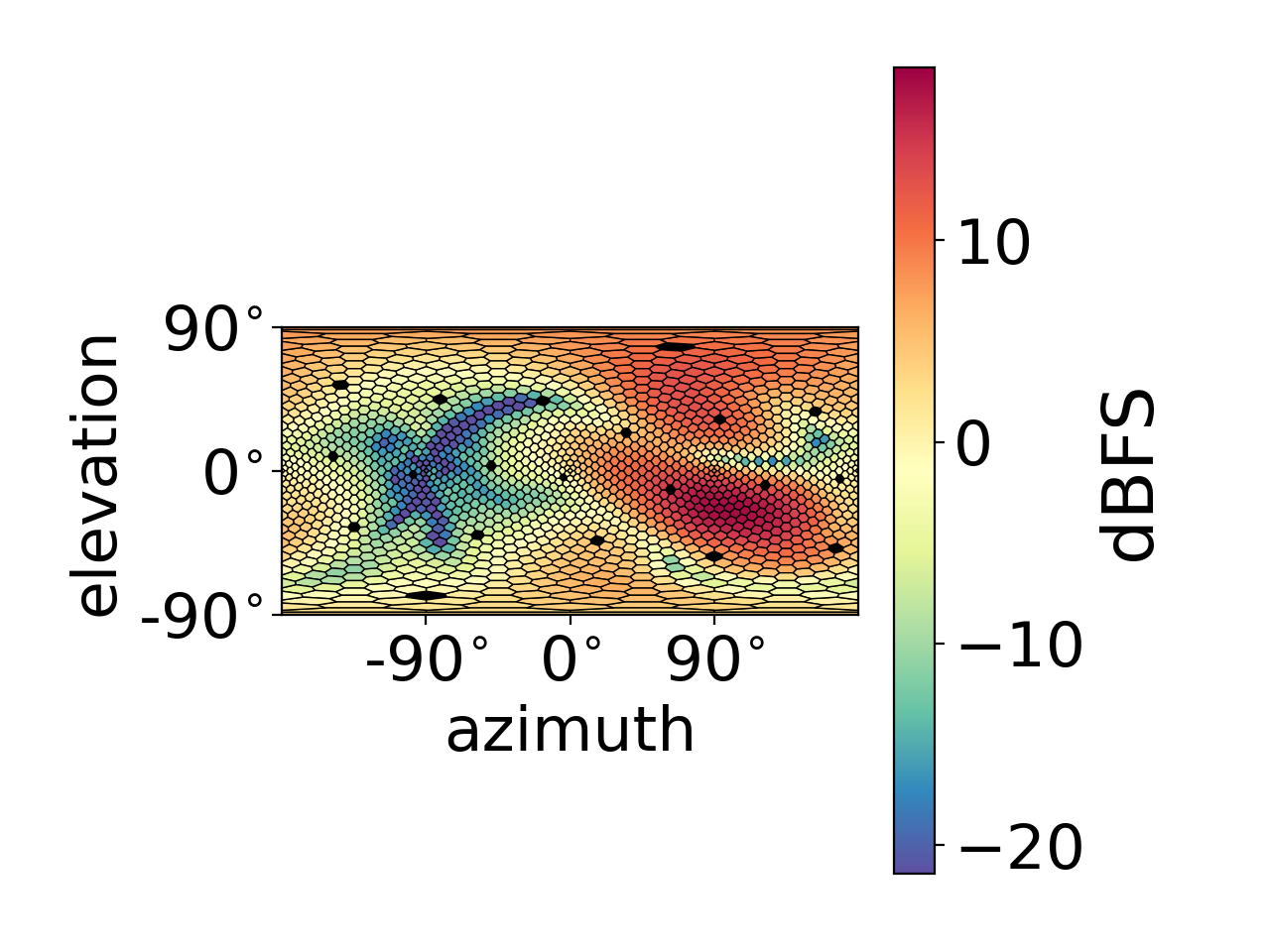}
		\end{tabular}
		\\
		\begin{tabular}[c]{@{}c@{}}
			$\qquad \angle{m_t}$\\
			\includegraphics[trim=18 30 0 70, clip, width=1.6in]{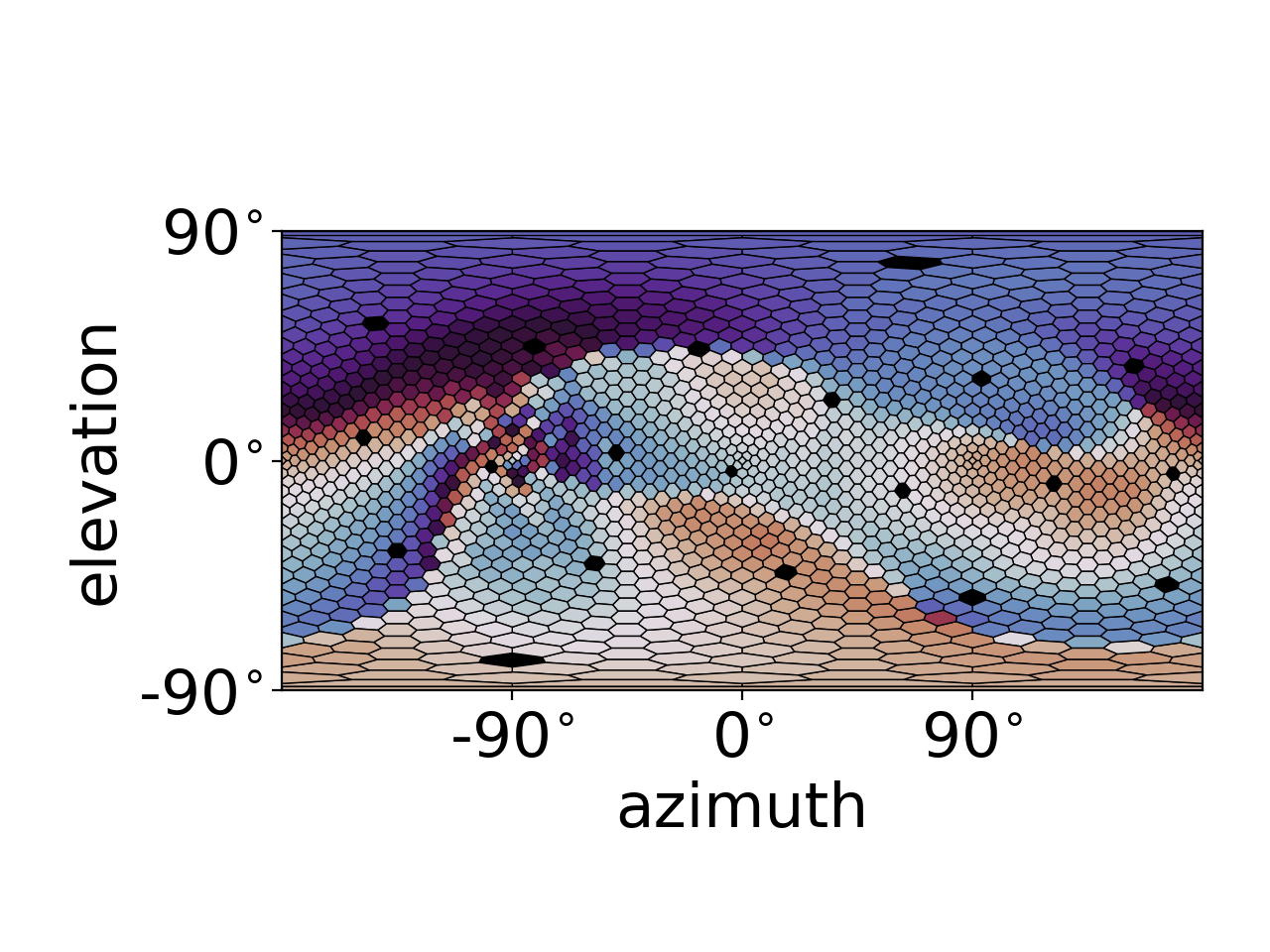}
		\end{tabular}
		&
		\begin{tabular}[c]{@{}c@{}}
			$\qquad \angle{\hat{m}_t}$\\
			\includegraphics[trim=18 30 0 70, clip, width=1.6in]{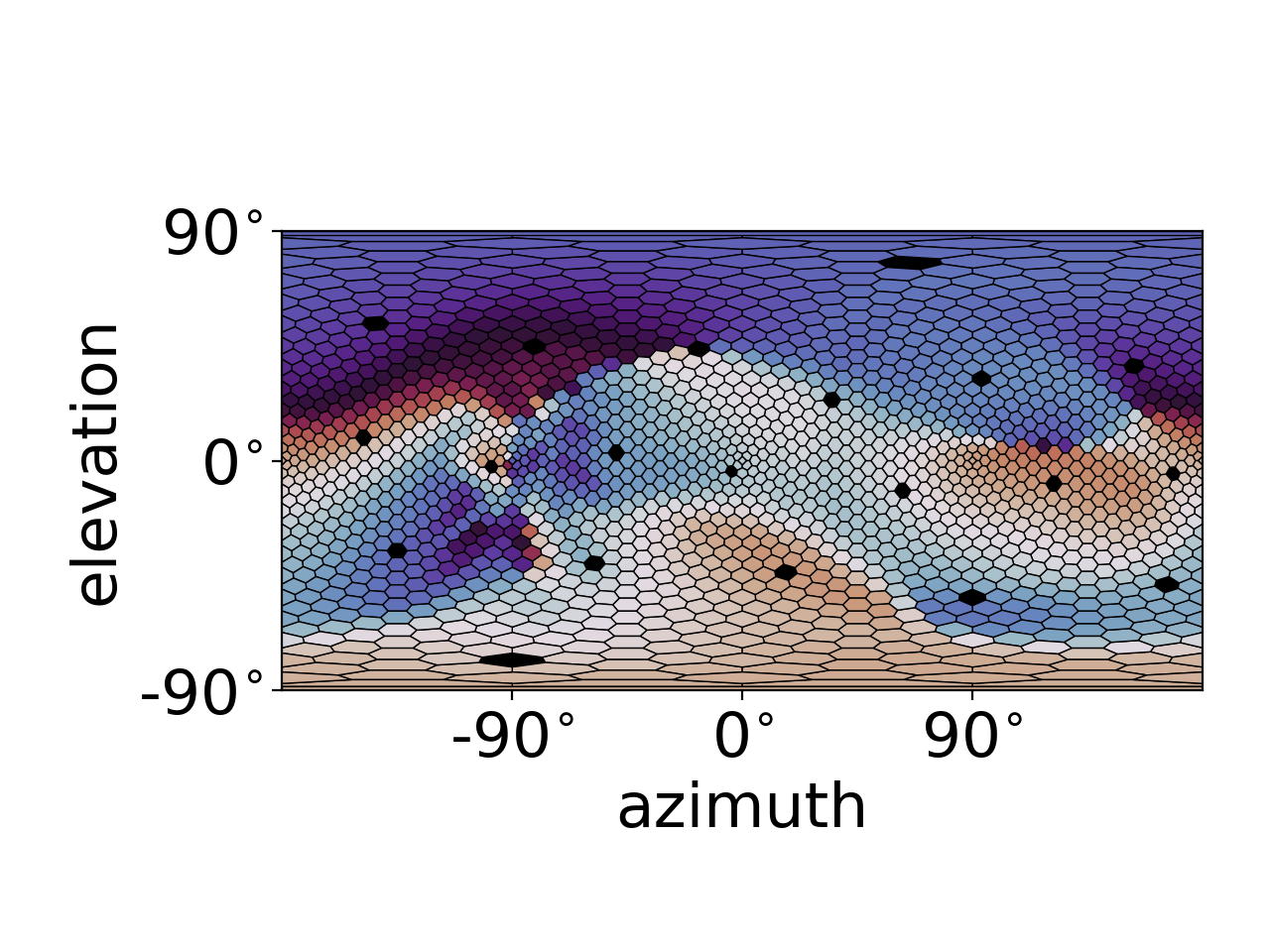}
		\end{tabular}
		&
		\begin{tabular}[c]{@{}c@{}}
			\includegraphics[trim=335 15 0 15, clip, width=0.4in]{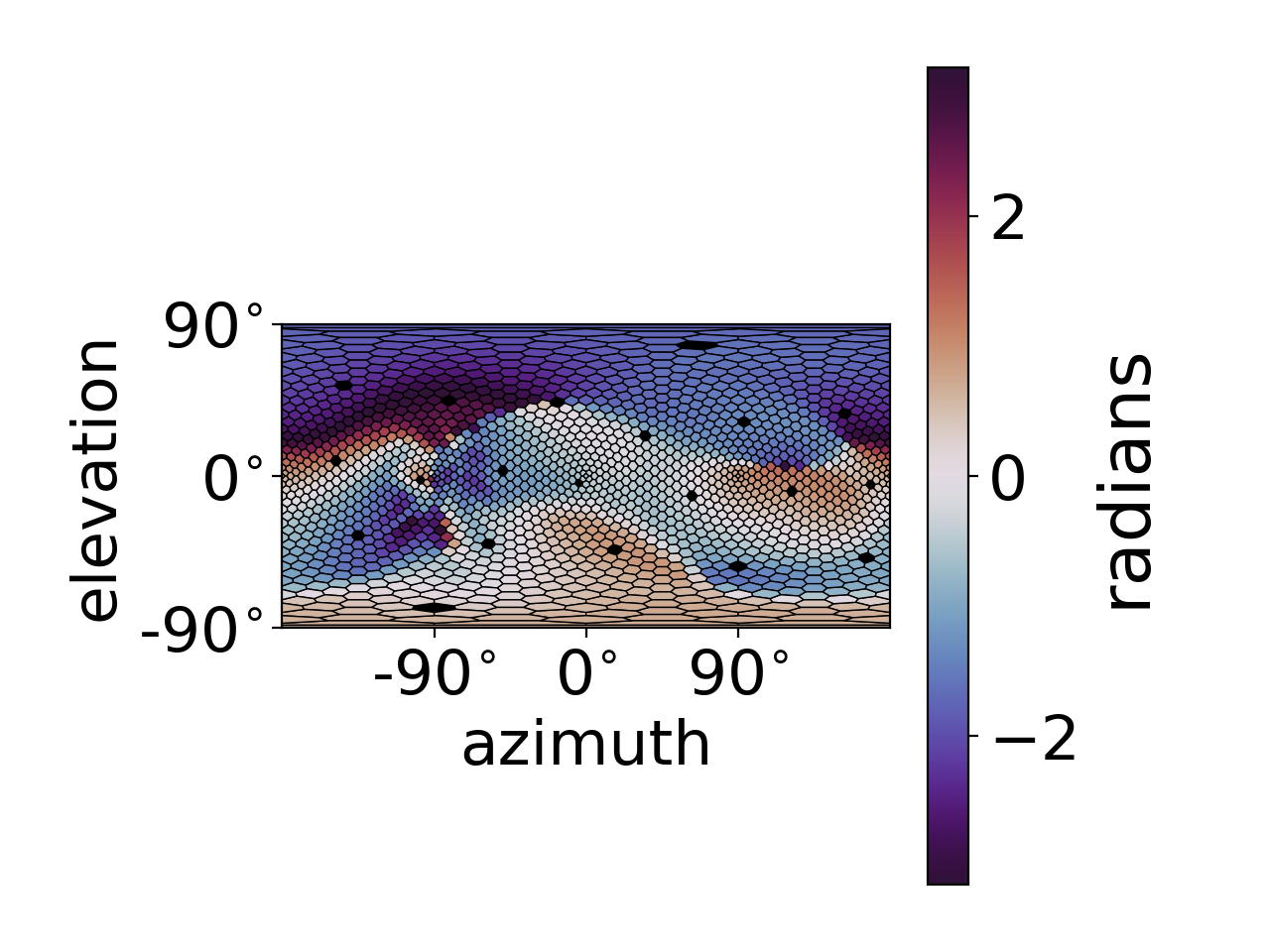}
		\end{tabular}
		\\
		\begin{tabular}[c]{@{}c@{}}
			$\qquad \left|\hat{m}_t-m_t\right|/\left|m_t\right|$\\
			\includegraphics[trim=18 30 0 70, clip, width=1.6in]{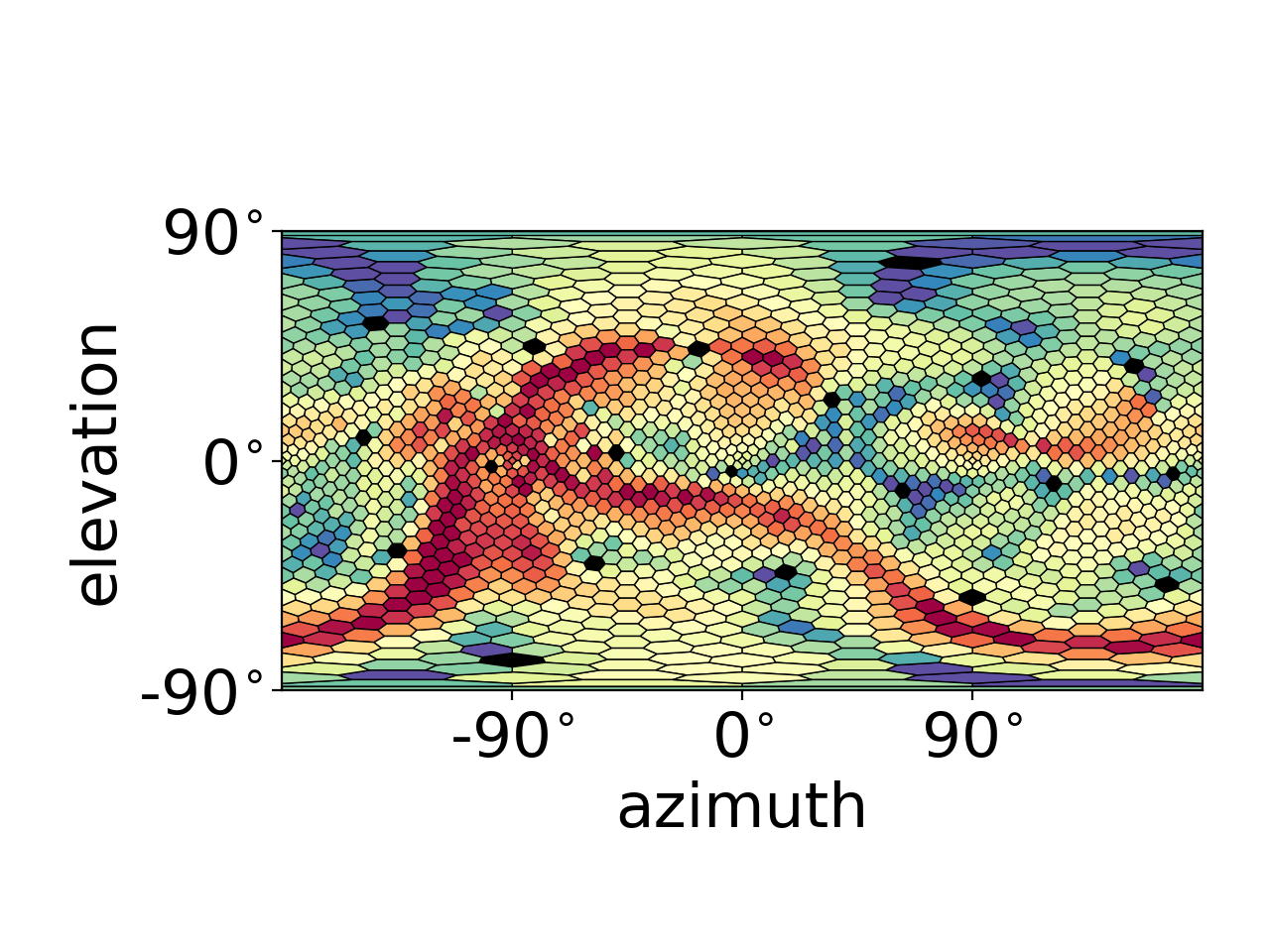}
		\end{tabular}
		&
		\begin{tabular}[c]{@{}c@{}}
			$\qquad \left|\sigma_t\right|/\left|m_t\right|$\\
			\includegraphics[trim=18 30 0 70, clip, width=1.6in]{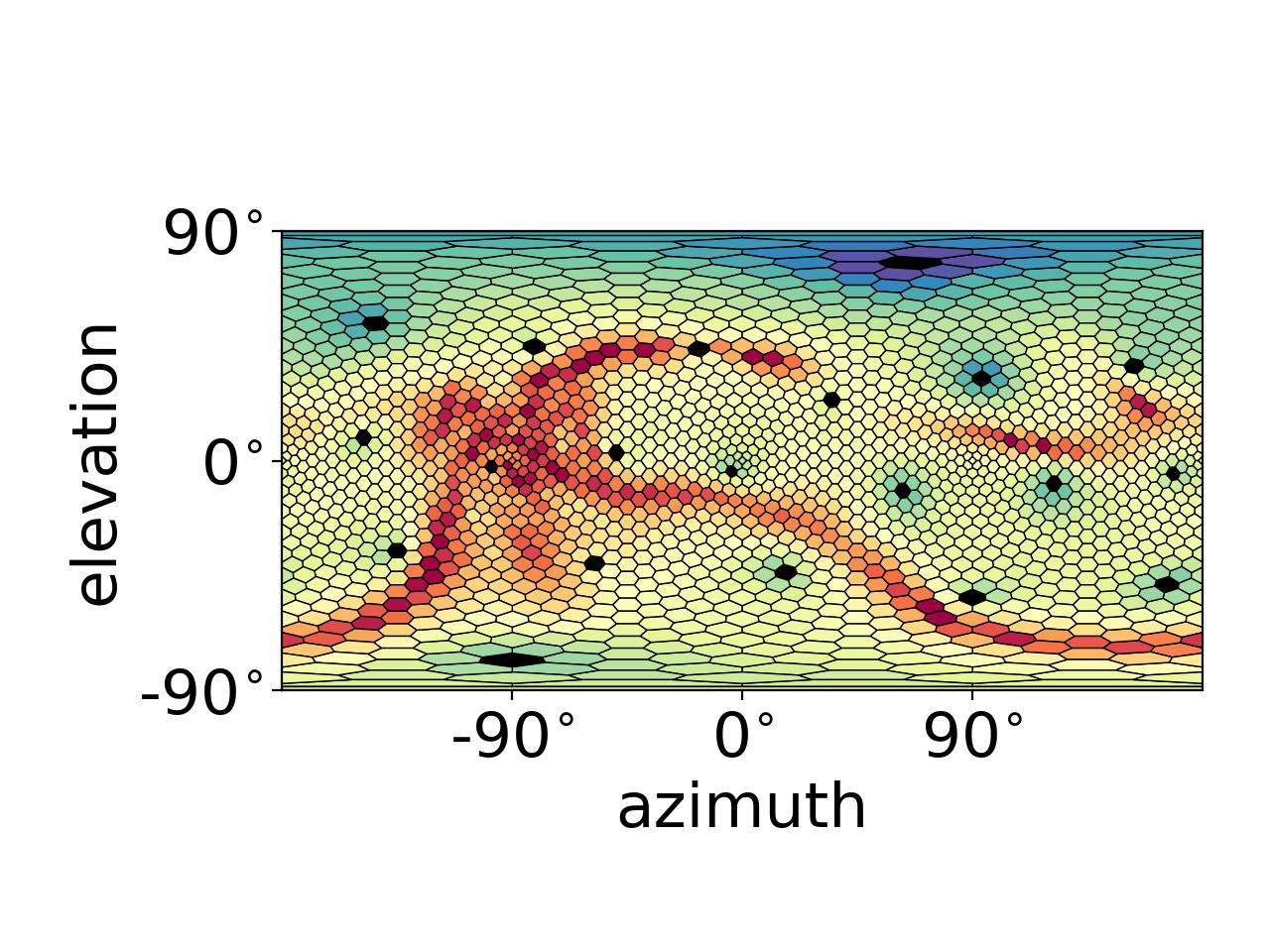}
		\end{tabular}
		&
		\begin{tabular}[c]{@{}c@{}}
			\includegraphics[trim=320 15 0 15, clip, width=0.4in]{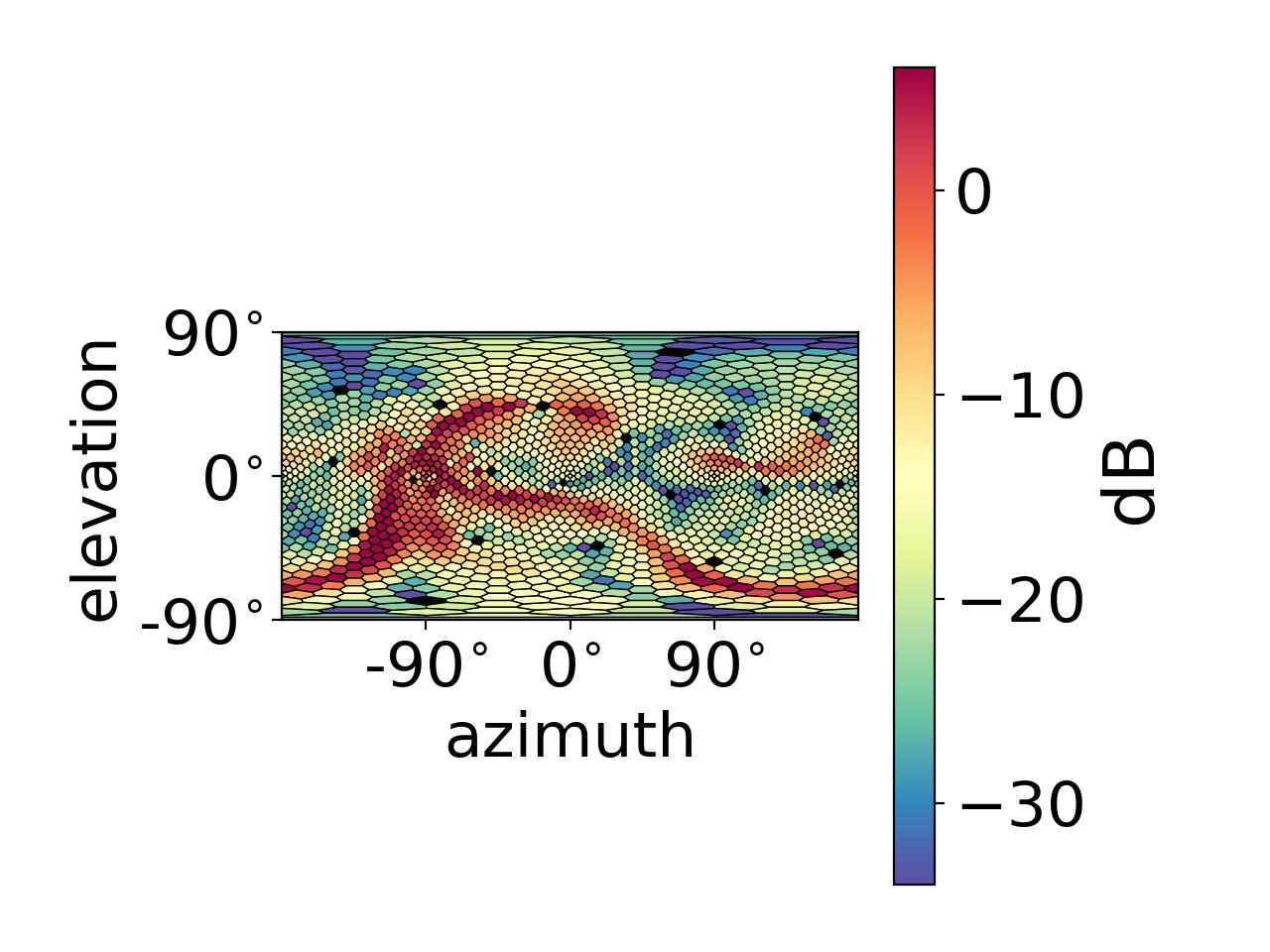}
		\end{tabular}
	\end{tabular}
\end{tabular}
\end{tabular}
\caption{Left ear channel time-aligned HRTF spectrum as a function of sound source direction $x_t$ for the interpolation task of Fig.~\ref{fig:log_magnitude_spectrum_example}. Plots are provided for the 58$^\text{th}$ frequency bin of the spectrum. Top and middle: log-magnitude and phase for ground truth $m_t=y_t+\bar{m_t}$ (left) and SConvCNP model's predictive mean $\hat{m_t}=\mu_t+\bar{m_t}$ (right) of the time-aligned HRTF spectrum. Bottom left: relative error committed by the SConvCNP model. Bottom right: SConvCNP predictive uncertainty relative to ground truth magnitude in decibels.\label{fig:sconvcnp_vs_ground_truth}}
\end{figure}

 The magnitude and phase responses of time-aligned HRTF spectrum is represented on HUTUB's data point grid for the 58$^\text{th}$ frequency bin in Fig.~\ref{fig:sconvcnp_vs_ground_truth}. As pictured, the SConvCNP model's mean estimate $\hat{m}_t=\mu_t+\bar{m}_t$ closely matches the ground truth $m_t=y_t+\bar{m}_t$ both in terms of magnitude and phase (top and middle plots). More precisely, the predictive mean solution's error generally lies under the -15 dB threshold relative to ground truth outside low-magnitude areas on the unit sphere (lower left plot). Furthermore, the predictive uncertainty provided by the model seems generally consistent with the observed error (lower right plot).

\par
\begin{figure}[t!]
\centering
\begin{tabular}{@{}c@{}c@{}}
\begin{tabular}[c]{@{}c@{}}
	\includegraphics[trim=22 22 22 22, clip, width=1.75in]{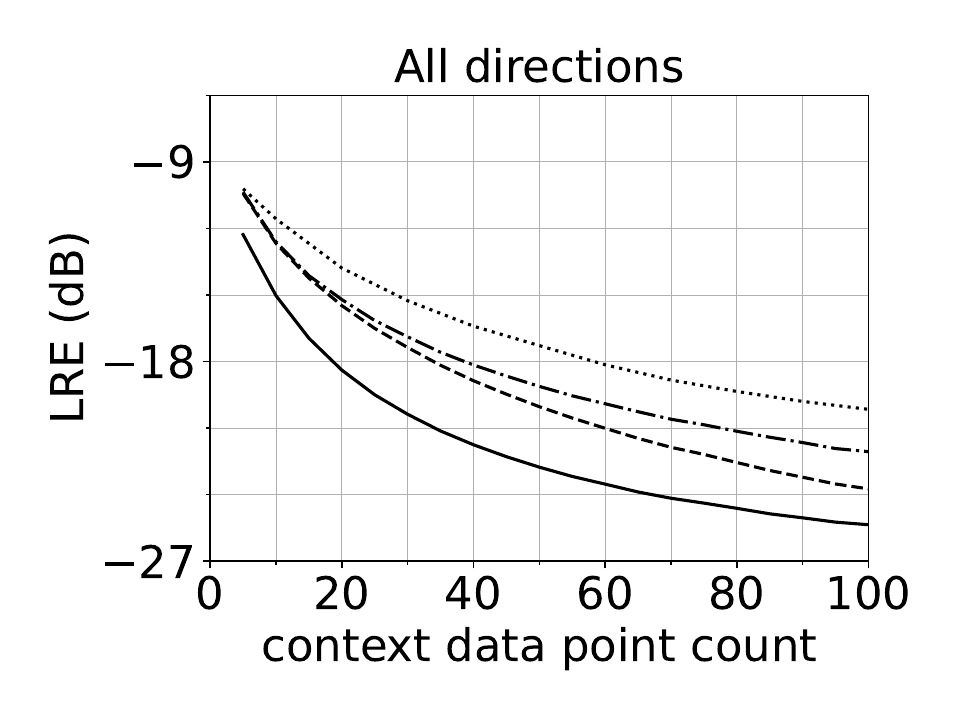}
\end{tabular}
&
\begin{tabular}[c]{@{}c@{}}
	\includegraphics[trim=22 22 22 22, clip, width=1.75in]{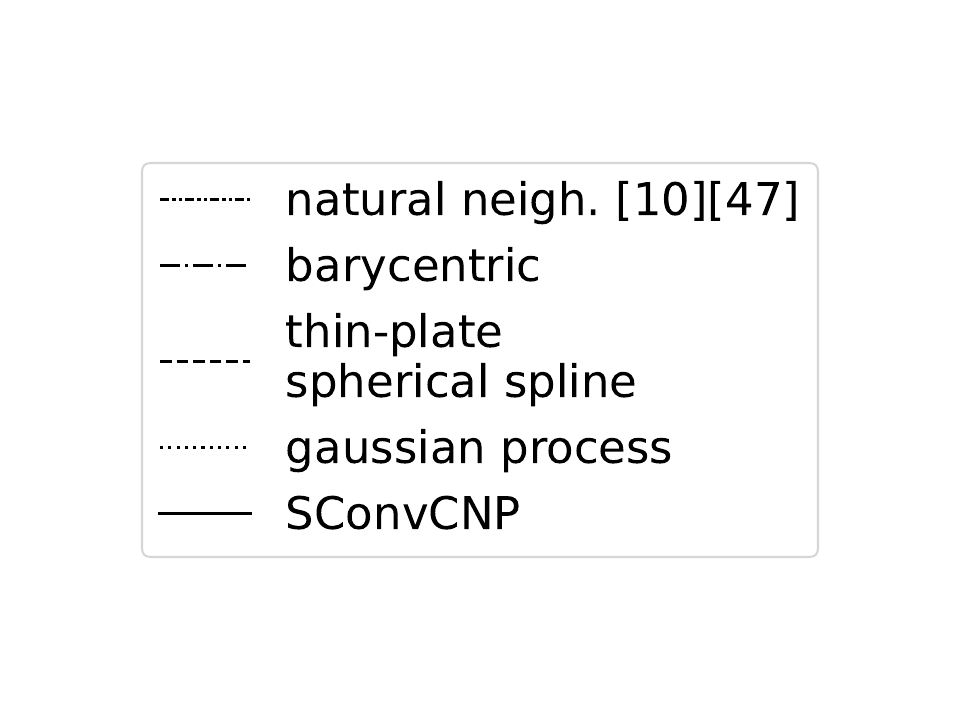}
\end{tabular}
\\
\begin{tabular}[c]{@{}c@{}}
	\includegraphics[trim=22 22 22 22, clip, width=1.75in]{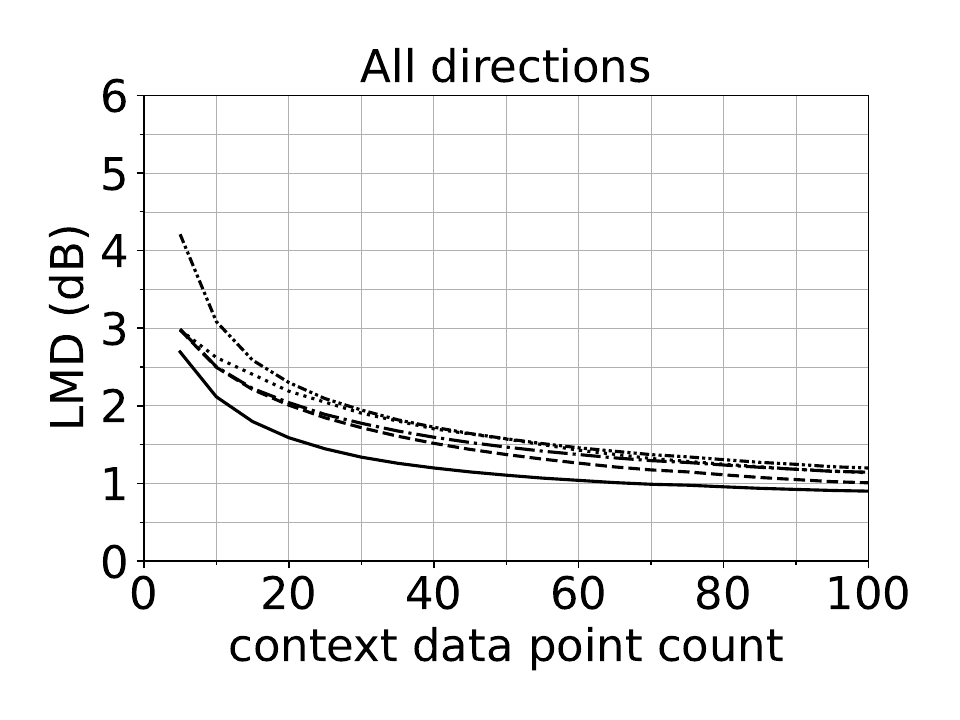}
\end{tabular}
&
\begin{tabular}[c]{@{}c@{}}
	\includegraphics[trim=22 22 22 22, clip, width=1.75in]{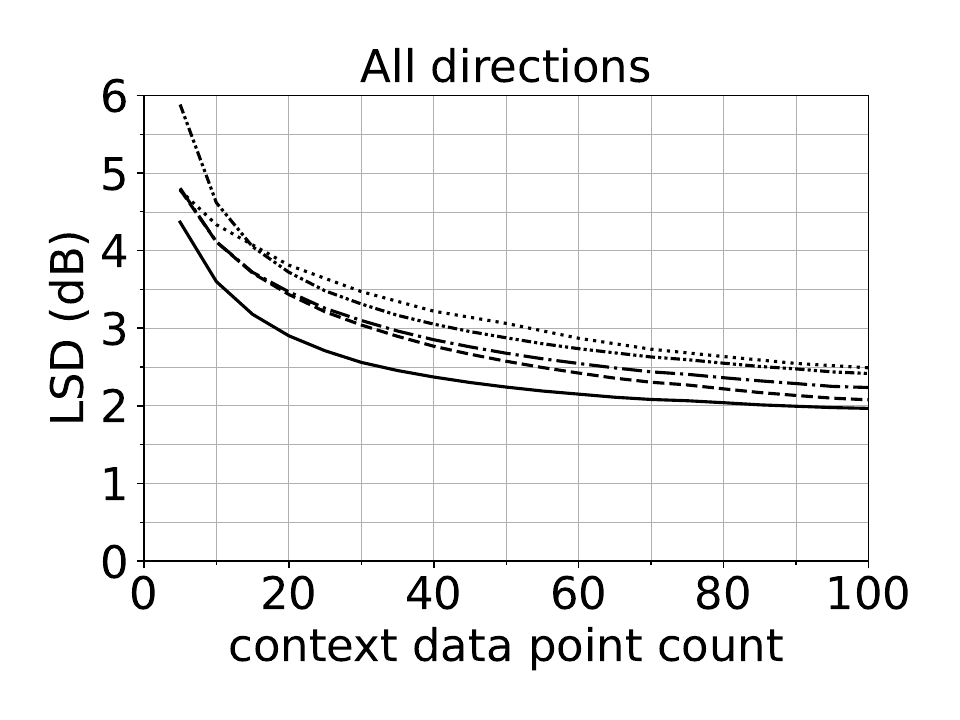}
\end{tabular}
\end{tabular}
\caption{Average time-aligned HRTF spectrum interpolation error across output features in the 0-15.5 kHz range and meta-test set's interpolation tasks as a function of context data point count. The proposed method (SConvCNP) improves upon all baselines on all three evaluation metrics. Upper-left: relative error level according to (\ref{eq:relative_error}). Lower-left: log-magnitude distance according to (\ref{eq:log_magnitude_distance}). Lower-right: log-spectral distortion according to (\ref{eq:log_spectral_distortion}).~\label{fig:sample_count_vs_metrics}}
\end{figure}

\begin{figure}[t!]

\centering
\begin{tabular}{@{}c@{}c@{}}
	\begin{tabular}[c]{@{}c@{}}
	\includegraphics[trim=0 40 0 40, clip, width=2.6in]{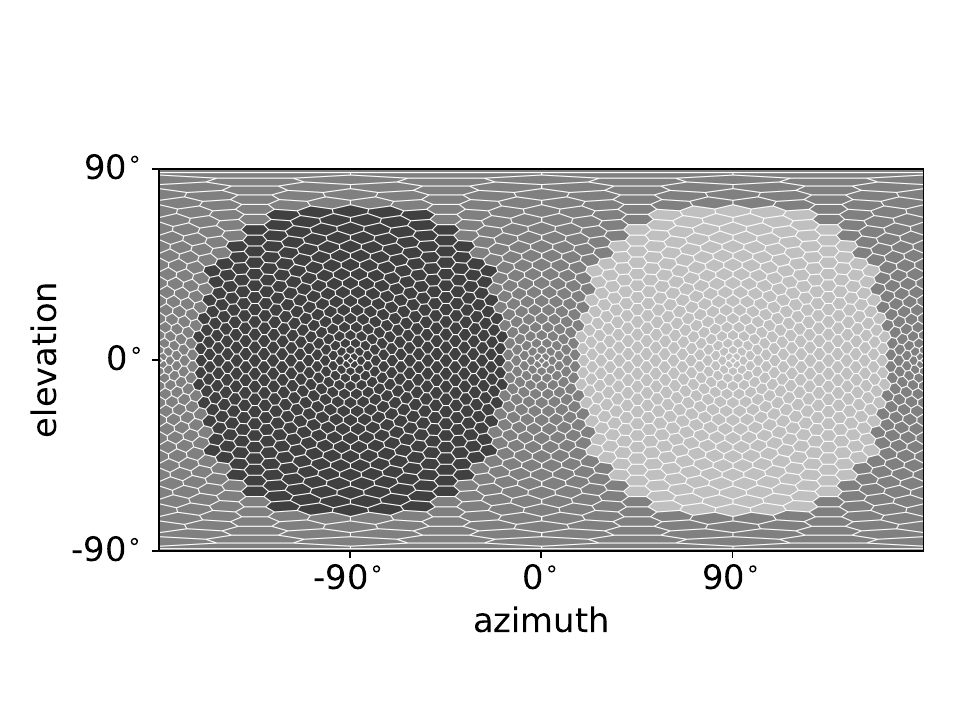}
	\end{tabular}
	&
	\begin{tabular}[c]{@{}c@{}}
	\includegraphics[trim=95 50 80 50, clip, width=0.7in]{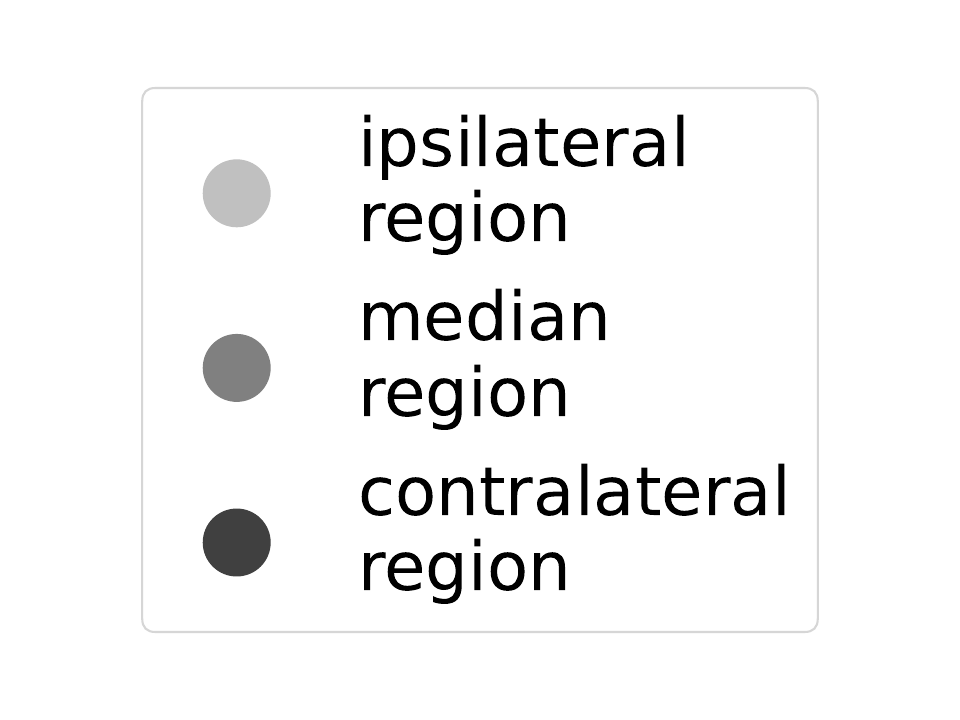}
	\end{tabular}
\end{tabular}
\caption{Definition of HRTF regions used in generating the plots of Fig.~\ref{fig:sample_count_vs_relative_error} and \ref{fig:frequency_vs_log_magnitude_error}. The boundaries separating the regions lies at $\pm 18.1^\circ$ lateral angle from the median plane, which distributes the HRTF directions of the HUTUBS grid in approximately equal proportions amongst the three specified regions. \label{fig:hrtf_regions}}
\end{figure}

\begin{figure}[t!]
\centering
\begin{tabular}{@{}c@{}c@{}}
\includegraphics[trim=22 22 22 22, clip, width=1.75in]{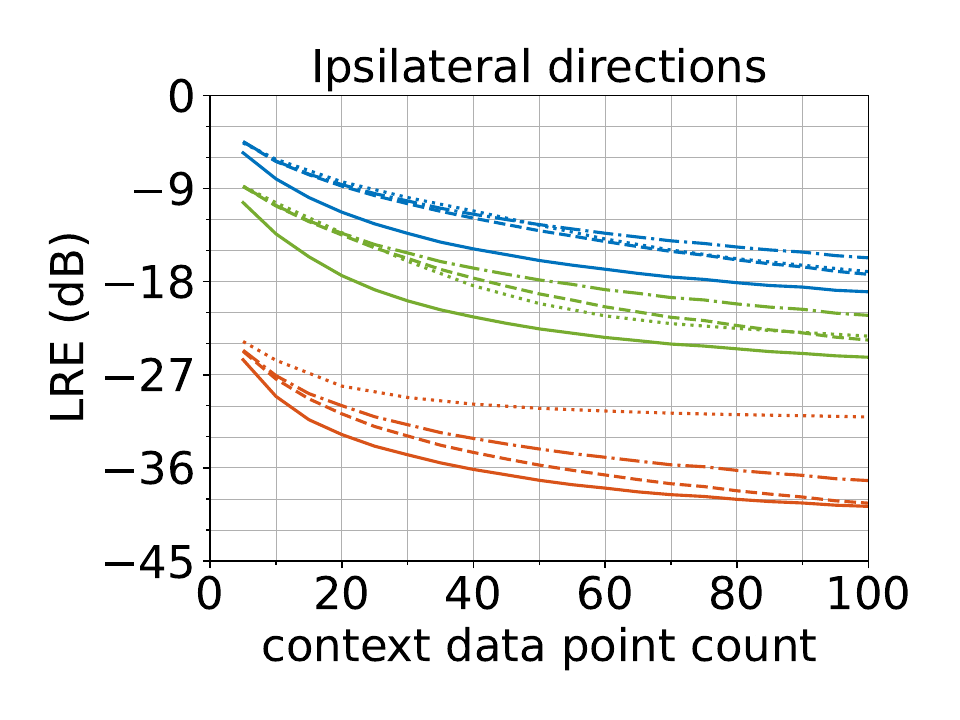}
&
\includegraphics[trim=22 22 22 22, clip, width=1.75in]{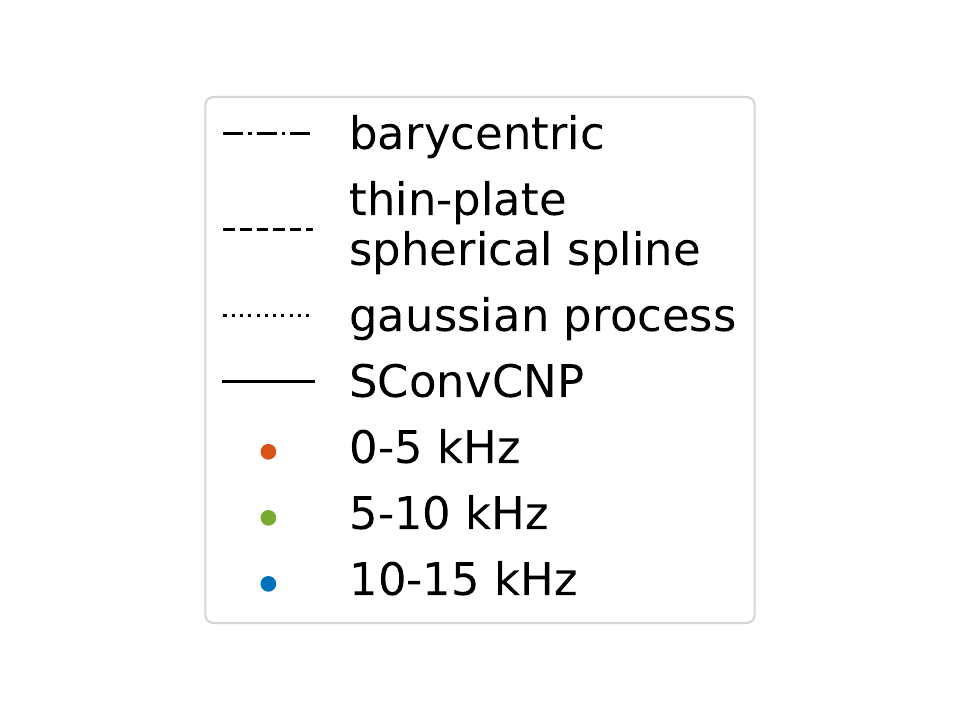}
\\
\includegraphics[trim=22 22 22 22, clip, width=1.75in]{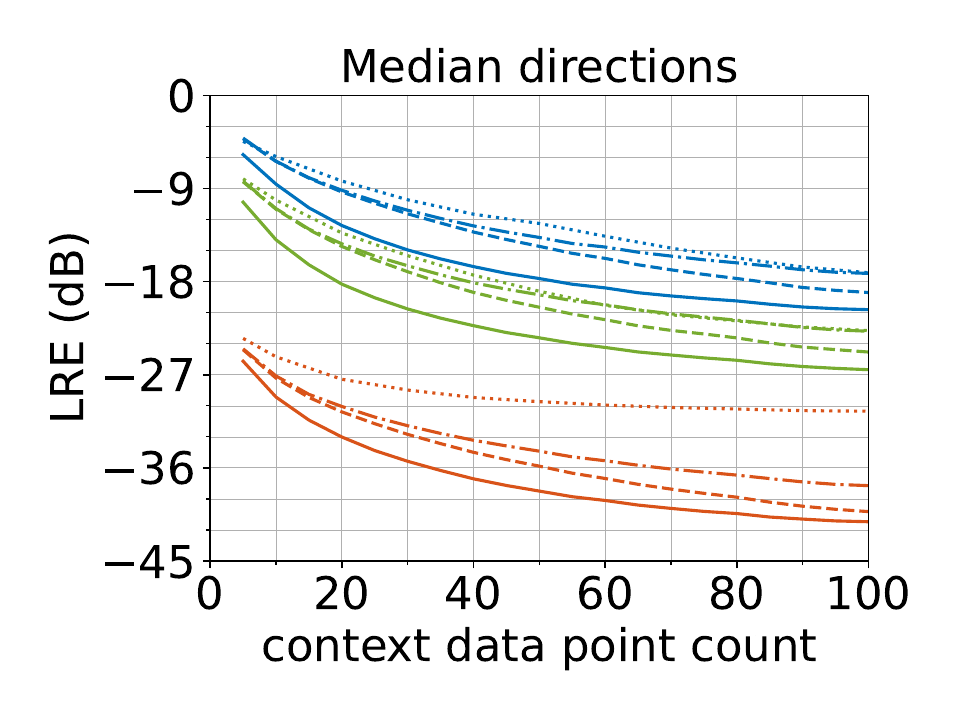}
&
\includegraphics[trim=22 22 22 22, clip, width=1.75in]{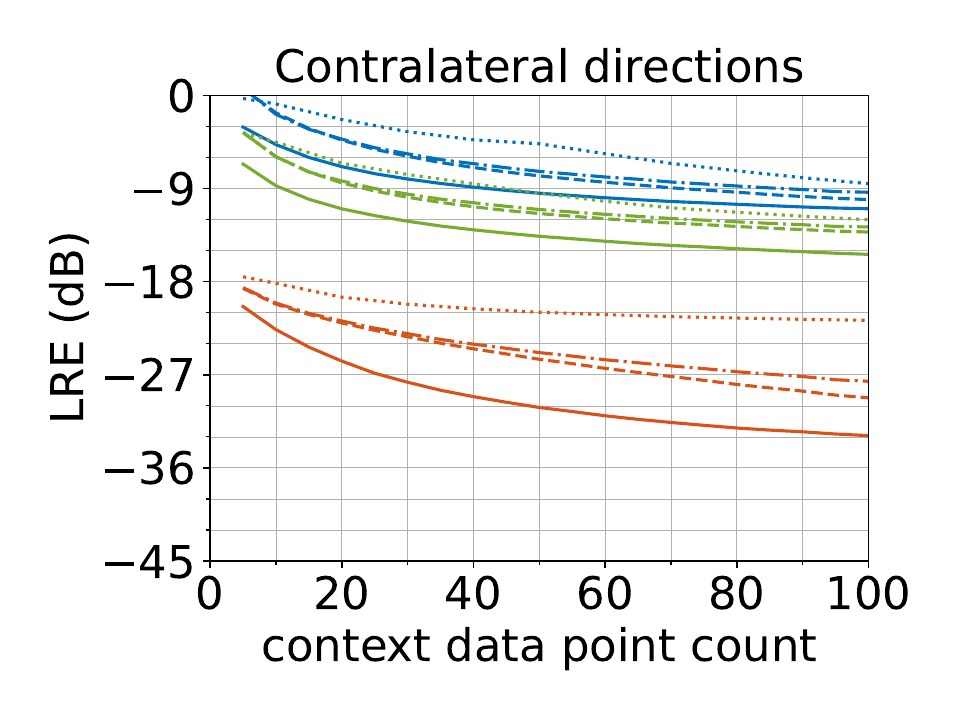}
\end{tabular}
\caption{Per frequency-band average time-aligned HRTF spectrum interpolation error across output features and the meta-test set's interpolation tasks as a function of context data point count. The proposed method (SConvCNP) improves upon all baselines in all specified regions and in each frequency-band. Upper-left, lower-left and lower-right: relative error level according to (\ref{eq:relative_error}) for HRTF direction sub-regions defined in Fig.~\ref{fig:hrtf_regions}.\label{fig:sample_count_vs_relative_error}}
\end{figure}

 A sample efficiency comparison of candidate methods is provided in Fig. \ref{fig:sample_count_vs_metrics} and \ref{fig:sample_count_vs_relative_error}. In particular, the graphs of these figures report error scores as a function of the number of context data points provided to the interpolation method candidates. Fig. \ref{fig:sample_count_vs_metrics} includes plots for the LRE, LMD and LSD metrics as defined in (\ref{eq:relative_error}), (\ref{eq:log_magnitude_distance}) and (\ref{eq:log_spectral_distortion}) respectively. Fig. \ref{fig:sample_count_vs_relative_error} provides further detail for the LRE metric specifically. In this figure, three additional LRE plots are provided for the ipsilateral, median, and contralateral HRTF regions defined in Fig.~\ref{fig:hrtf_regions}. Error levels are provided in each plot of Fig.~ \ref{fig:sample_count_vs_relative_error} for three distinct frequency bands: 0-5 kHz, 5-10 kHz, and 10-15 kHz. The natural neighbor method is intentionally omitted from LRE plots of Figs~\ref{fig:sample_count_vs_metrics} and \ref{fig:sample_count_vs_relative_error} as this candidate can only be meaningfully compared on magnitude-error-metric grounds since it interpolates the HRTF spectrum without the time-alignement pre-processing. In both Figs. \ref{fig:sample_count_vs_metrics} and \ref{fig:sample_count_vs_relative_error}, each curve represents an average error score value taken across tasks, directions, left/right ear channels and, the case being, frequency bins. In particular, the average was taken for each reported count over a set of 340 randomly drawn meta-test tasks.

\par As expected, all candidates exhibit monotonically decreasing error scores with increased sample count. This observation holds for all error metrics considered in Fig. \ref{fig:sample_count_vs_metrics} and for all regions and frequency intervals considered in Fig. \ref{fig:sample_count_vs_relative_error}. Moreover, the SConvCNP model presents significantly lower relative error level compared to the thin-plate spherical spline method, which forms the best baseline: up to 3 dB globally (top left plot of Fig. \ref{fig:sample_count_vs_metrics}) and up to 4.5 dB in the 0-5 kHz range (contralateral region, bottom right plot of Fig. \ref{fig:sample_count_vs_relative_error}). This improvement translates to nearly a halving of required measurement count to meet an error specification level. For example, meeting an -20 dB average relative error requires approximately 50 measurements using the thin-plate spherical spline method while approximately 28 is sufficient on average using the SConvCNP model. Similar observations can be made in the case of both the LMD and LSD metrics pictured in Fig. \ref{fig:sample_count_vs_metrics}.

\par
\begin{figure}[t!]
\centering
\begin{tabular}{@{}c@{}c@{}}
	\begin{tabular}[c]{@{}c@{}}
		\includegraphics[trim=22 22 22 22, clip, width=1.75in]{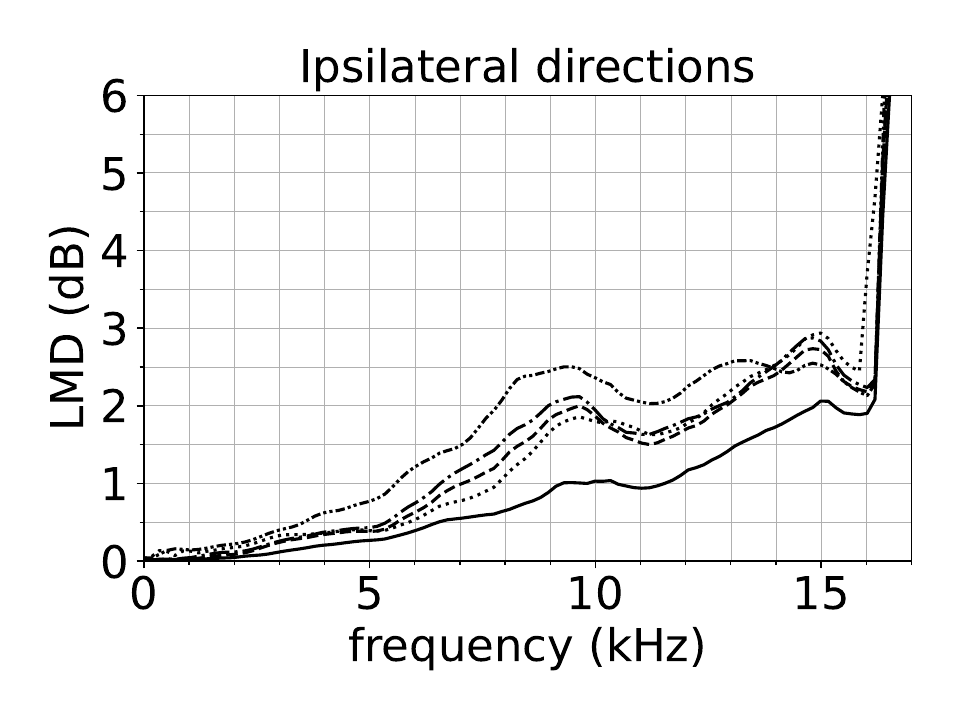}
	\end{tabular}
&
	\begin{tabular}[c]{@{}c@{}}
		\includegraphics[trim=22 22 22 22, clip, width=1.75in]{./figures/all_directions_plot_legend.pdf}
	\end{tabular}
 \\
	\begin{tabular}[c]{@{}c@{}}
		\includegraphics[trim=22 22 22 22, clip, width=1.75in]{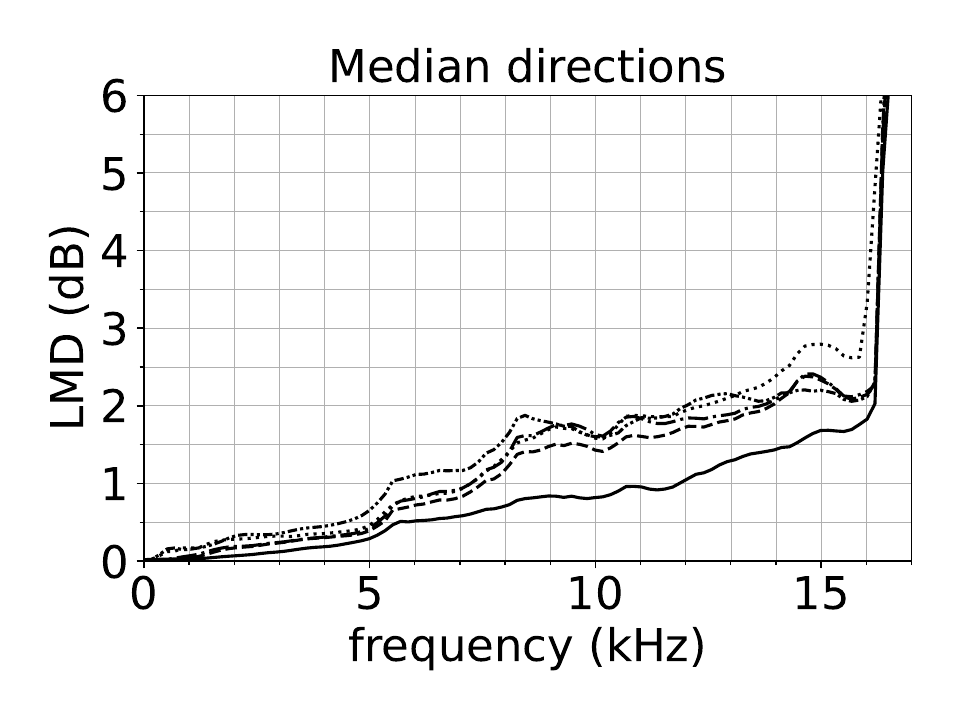}
	\end{tabular}
&
	\begin{tabular}[c]{@{}c@{}}
		\includegraphics[trim=22 22 22 22, clip, width=1.75in]{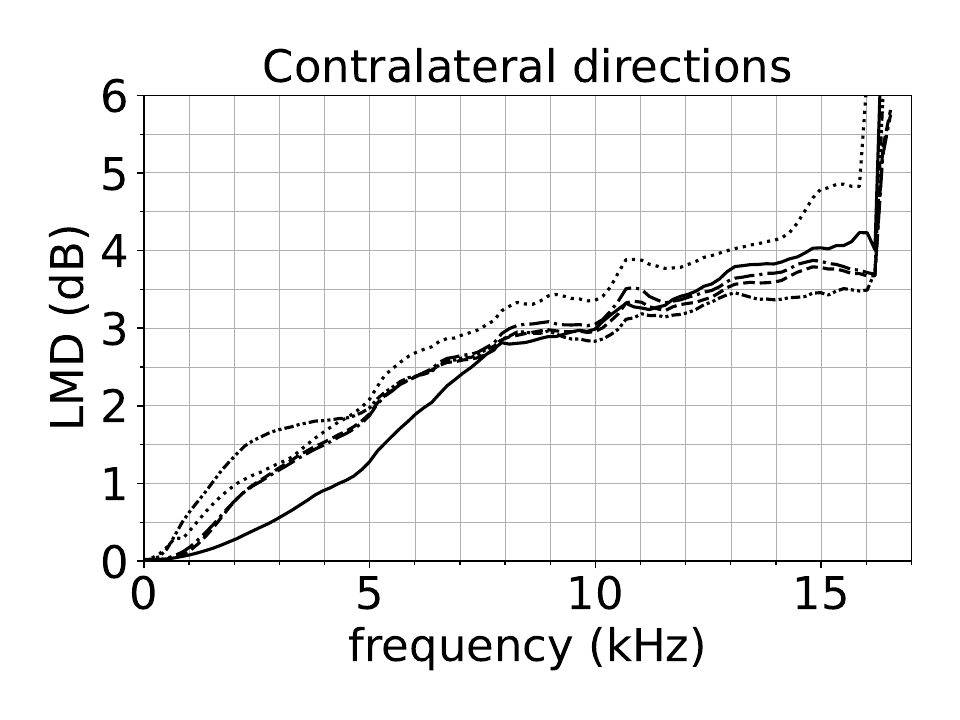}
	\end{tabular}
\end{tabular}
\caption{Average log-magnitude distance as a function of frequency for meta-test tasks with numbering 40 context data points. Each plot corresponds to one of the three HRTF direction regions defined in Fig.~\ref{fig:hrtf_regions}.\label{fig:frequency_vs_log_magnitude_error}}
\end{figure}

 Fig. \ref{fig:frequency_vs_log_magnitude_error} provides a summary of log-magnitude distance level as a function of frequency in the specific case of context sets numbering 40 context data points. The three plots of the figure detail the error levels specific to each region defined in Fig.~\ref{fig:hrtf_regions}. %
The SConvCNP candidate significantly outperforms all baselines in the 0-14 kHz across all regions, except in the contralateral region (top-right plot) where the natural neighbour matches and then outperforms the proposed model from the 7.5 kHz mark onwards. In agreement with the results of Fig. \ref{fig:sample_count_vs_relative_error}, the improvements brought by the SConvCNP model are most significant beyond the 6 kHz mark in the frontal and ipsilateral region, while it is most significant under 6 kHz in the contralateral region. In particular, the SConvCNP model provides an improvement of up to 0.8 dB compared to the best baseline at any frequency, as found in the ipsilateral region around the 9.2 kHz mark.%

\par
\begin{figure}[t!]
\centering
\begin{tabular}{@{}c@{}c@{}}
\includegraphics[width=1.77in]{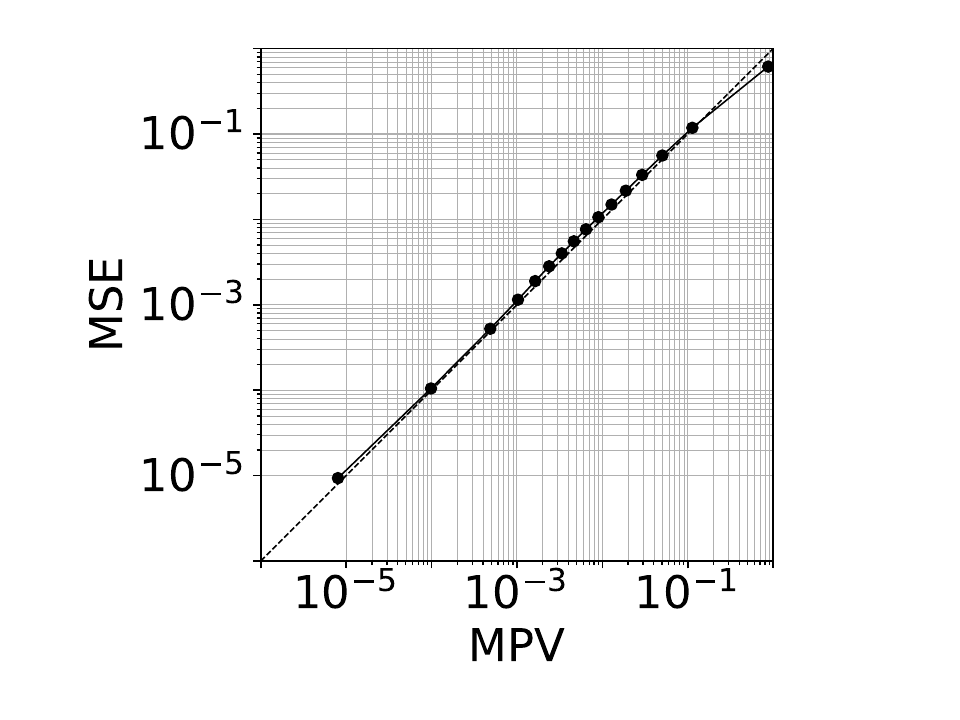} &
\includegraphics[width=1.77in]{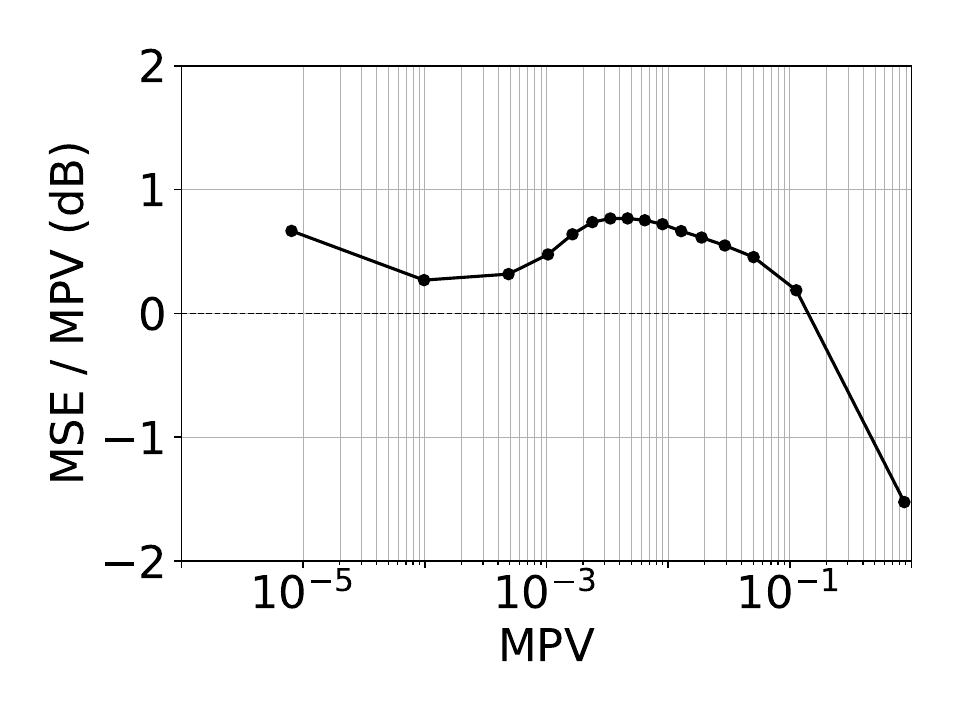}
\end{tabular}
\caption{Miscalibration of the trained SConvCNP model's uncertainty estimates. Left: mean predicted variance (horizontal axis) versus mean square error (vertical axis) plot. Perfect calibration corresponds to the identity line (dashed). Right: miscalibration level in decibels.\label{fig:rmv_vs_rmse}}
\end{figure}

 Miscalibration of the trained SConvCNP model is summarized in Fig. \ref{fig:rmv_vs_rmse}. In this figure, the calibration of the trained SConvCNP model is evaluated over a meta-test set of 340 randomly generated tasks using $D=16$ divisions. In particular, the predicted variance and squared error pairs observed at the output the model are pooled across interpolation tasks, data point locations, frequency bins, left/right channels, and real/imaginary parts to form the MSE versus MPV curve shown in Fig. \ref{fig:rmv_vs_rmse}.

\par As pictured in the left plot of Fig.~\ref{fig:rmv_vs_rmse}, the mean predicted variance closely matches the mean square error. In particular, the resulting curve lies neither significantly above or below the identity line. Hence, the trained SConvCNP model is neither markedly over-confident or under-confident in its predictions. More precisely, the rightmost plot reveals that the effective squared error lies, in expectation and for all but most uncertain predictions, within 1.0 dB of the predicted variance. This level of miscalibration is moderate when put in contrast to the range of $\sim$9 dB relative error reduction that is achieved when acquiring data points from a count of 5 to a count of 40 as pictured in the top-left plot of Fig.~\ref{fig:sample_count_vs_metrics}. %
Accordingly, we conclude that the model's uncertainty estimates $\sigma_t$ can usefully inform the problem of acquiring additional HRTF data points to improve HRTF individualization upon a pre-existing HRTF estimate.

%% file: sections/conclusion.tex
\par In this work we introduced the first HRTF interpolation method providing well-calibrated uncertainty estimates. We showed the method proved sample efficient on the time-aligned HRTF spectrum interpolation task. In particular, meta-training was carried-out successfully using a modest data set of 85 subjects.
Furthemore, the interpolators returned by the proposed meta-learning model were shown to require up to nearly half the number of context data point count compared to state-of-the-art interpolation methods at comparable accuracy level. Contrary to the Gaussian process regression baseline, they also showed well-calibrated uncertainty estimates.
\par The proposed model's  time and space complexity severely limits its applicability towards real-time interpolation and audio rendering setups. However it can readily be used for offline up-sampling of sparse HRTF sets. 
Furthermore, a promising application lies in facilitating the sequential decision problem of acquiring as few correcting HRTF data-points as needed to achieve a required degree of HRTF individualization accuracy. In particular, the provided uncertainty estimates could be used to identify the location at which obtaining a new measurement would, in expectation, maximally reduce the model's uncertainty.  Furthermore, the predictive distribution could be used to compare the probability of HRTF query candidates conditioned on the data points already acquired for the subject so as to select the most relevant ones to be submitted for perceptual feedback evaluation from the subject. %
Treatment of such sequential decision problem is left as future research work. 
Other future development avenues include evaluation of the model's ability to correct HRTF estimates provided by state-of-the-art parametric individualization methods instead of the train set population mean used under the limited scope of this work. Finally, we intend to prepare and provide publicly-available versions of the model and code.